\documentclass[10pt]{revtex4}
\usepackage{graphicx}

\usepackage[american]{babel}
\usepackage{amssymb}
\usepackage{amsfonts}
\usepackage{color}
\begin{document}
\newtheorem{corollary}{Corollary}[section]
\newtheorem{remark}{Remark}[section]
\newtheorem{definition}{Definition}[section]
\newtheorem{theorem}{Theorem}[section]
\newtheorem{proposition}{Proposition}[section]
\newtheorem{lemma}{Lemma}[section]
\newtheorem{help1}{Example}[section]
\renewcommand{\theequation}{\arabic{section}.\arabic{equation}}

\title{Breathers for the Discrete Nonlinear Schr{\"o}dinger equation with nonlinear hopping}
\author{N. I. Karachalios$^1$, B. S\'{a}nchez-Rey$^2$, P. G. Kevrekidis$^3$ and J. Cuevas$^2$}

\affiliation{
$^1$ Department of Mathematics, University of the Aegean,\\
  Karlovassi, 83200 Samos, Greece\\
$^2$ Grupo de F\'{\i}sica No Lineal. Departamento de F\'{\i}sica Aplicada I, Universidad de Sevilla, Escuela
    Polit\'{e}nica Superior,\\
    C/ Virgen de Africa, 7, University of Sevilla,\\
    41011 Sevilla, Spain\\
$^3$ Department of Mathematics and Statistics, University of Massachusetts\\
  Lederle Graduate Research Tower,\\
  Amherst MA 01003-9305, USA \\}
\date{\today}
\begin{abstract}
We discuss the existence of breathers and lower bounds on their
power, in nonlinear Schr{\"o}dinger lattices with nonlinear hopping.
Our methods extend from a simple variational approach to fixed point arguments,
deriving lower bounds for the power which can serve as a threshold for the
existence of breather solutions. Qualitatively, the theoretical results
justify non-existence of breathers below the prescribed lower bounds
of the power
%independently of
which depend on the dimension, the parameters of the lattice
%, although quantitatively, the
%lower bounds are functions of these  parameters
as well as of the frequency of
breathers. In the case of supercritical power nonlinearities we investigate the interplay of these estimates with the optimal constant of the discrete interpolation inequality. Improvements of the general estimates, taking into account the localization of the true breather solutions are derived.
%Extended
Numerical studies in the one dimensional lattice
corroborate the theoretical bounds
%can be considered as thresholds for the
%existence of breathers,
and illustrate
that in certain parameter regimes of physical significance, the estimates
can serve as accurate predictors of the breather power and its
dependence on the various system parameters.
%energy thresholds for breather existence.
\end{abstract}
\maketitle
\section{Introduction}

The discrete nonlinear Schr{\"o}dinger (DNLS) model constitutes a ubiquitous
example of a nonlinear dynamical lattice with a wide range of
applications, extending from the nonlinear optics of fabricated AlGaAs
waveguide arrays as in \cite{dnc_review,kivshar_review,moti_review},
to the atomic physics
of Bose-Einstein condensates in sufficiently deep optical
lattices analyzed in \cite{bec_reviews,bec_reviewsA,bec_reviewsB,bec_reviewsC}.
Partly also due to these applications, the DNLS has been a focal
point of numerous mathematical/computational investigations in its
own right, a number of which has been summarized in \cite{dnls,Eil,Panos_book,reviewsA,reviewsB,reviewsC}
and is related to models used in numerous other settings
including micromechanical cantilever arrays \cite{sievers} and
DNA breathing dynamics \cite{peyrard}, among others.

In this work we consider a variant of the DNLS equation of the following
form:
\begin{eqnarray}
\label{hopDNLS} \mathrm{i}\dot{\psi}_n+\epsilon(\Delta_d\psi)_n+
\alpha\psi_n\sum_{j=1}^N(\mathcal{T}_{j}\psi)_{n\in\mathbb{Z}^N}
+\beta|\psi_n|^{2\sigma}\psi_n=0,
\end{eqnarray}
on a $N$-dimensional lattice which can be finite if supplemented with Dirichlet boundary conditions, or infinite ($n\in\mathbb{Z}^N$). In (\ref{hopDNLS}),  $\epsilon>0$ is a discretization parameter $\epsilon\sim h^{-2}$
with $h$ being the lattice spacing, and  $(\Delta_d\psi)_n$ stands for the $N$-dimensional discrete Laplacian
\begin{eqnarray}
\label{DiscLap}
(\Delta_d\psi)_{n\in\mathbb{Z}^N}=\sum_{m\in \mathcal{N}_n}\psi_m-2N\psi_n,
\end{eqnarray}
where $\mathcal{N}_n$ denotes the set of $2N$ nearest neighbors of the point in $\mathbb{Z}^N$ with label $n$.
The nonlinear operator $\mathcal{T}_{j}$ is defined for every $\psi_n$,
$n=(n_1,n_2,\ldots,n_N)\in\mathbb{Z}^N$, as
\begin{eqnarray}
 \label{defopT}
(\mathcal{T}_{j}\psi)_{n\in\mathbb{Z}^N}=|\psi_{(n_1,n_2,\ldots,n_{j}+1,n_{j+1},\ldots,n_{N})}|^2+|\psi_{(n_1,n_2,\ldots,n_{j}-1,n_{j+1},\ldots,n_{N})}|^2,\;\;j=1,\ldots,N.
\end{eqnarray}
The nonlinearity parameters $\alpha,\beta\in\mathbb{R}$. In the case $\alpha=0$, $\beta\neq 0$, one recovers the classical DNLS equation with power nonlinearity. The case where $\alpha,\beta\neq 0$, corresponds to the DNLS equation with nonlinear hopping terms.
The DNLS equation (\ref{hopDNLS}), is a Hamiltonian model with a
Hamiltonian of the form:
\begin{eqnarray}
\label{Hamhop}
\mathcal{H}[\psi]=\epsilon(-\Delta_d\psi,\psi)_2
-\sum_{j=1}^N\sum_{n_j=-\infty}^{+\infty}|\psi_{(n_1,n_2,\ldots,n_{j},n_{j+1},\ldots,n_{N})}|^2|\psi_{(n_1,n_2,\ldots,n_{j}+1,n_{j+1},\ldots,n_{N})}|^2-\frac{\beta}{\sigma+1}\sum_{n\in\mathbb{Z}^N}
|\psi_n|^{2\sigma+2}.
\end{eqnarray}
Let us note for convenience discuss the $1-\mathrm{D}$ lattice, where
the equation (\ref{hopDNLS}) reads:
\begin{eqnarray}
\label{hopDNLS1D}
\mathrm{i}\dot{\psi}_n+\epsilon(\psi_{n-1}-2\psi_n+\psi_{n+1})+
\alpha\psi_n\left(|\psi_{n+1}|^2
+|\psi_{n-1}|^2\right)+\beta|\psi_n|^{2\sigma}\psi_n=0,
\end{eqnarray}
with the Hamiltonian
\begin{eqnarray}
\label{Hamhop1D}
\mathcal{H}[\psi]=\epsilon\sum_{n\in\mathbb{Z}}|\psi_{n+1}-\psi_{n}|^2
-\alpha\sum_{n\in\mathbb{Z}}|\psi_n|^2|\psi_{n+1}|^2-\frac{\beta}{\sigma+1}\sum_{n\in\mathbb{Z}}
|\psi_n|^{2\sigma+2}.
\end{eqnarray}
The Hamiltonian (\ref{Hamhop}) and the power (or norm)
\begin{eqnarray}
\label{power}
\mathcal{P}[\psi]=\sum_{n\in\mathbb{Z}^N}
|\psi_n|^{2}
\end{eqnarray}
are the conserved quantities of this lattice dynamical system.

We will present theoretical and numerical results related to the  existence of time periodic (standing wave) solutions of the form
\begin{eqnarray}
\label{TP}
\psi_n(t)&=&e^{\mathrm{i}\Omega
  t}\phi_n,\;\Omega\in\mathbb{R}.
\end{eqnarray}

The physical interest in this particular model stems from various contexts, as the modeling of quantum lattices and waveguide arrays and the approximation of the dynamics Klein-Gordon (KG) and Fermi-Pasta-Ulam (FPU) chains \cite{Falvo, claude, joha, joha2, joha3, joha4}.  Eq. (\ref{hopDNLS1D}) for cubic ($\sigma=1$) nonlinearity
corresponds to the classical limit of the quantum DNLS equation introduced
in \cite{Falvo}. In the quantum lattice introduced therein, the inclusion of
the nonlinear hopping term allows a fast energy propagation as long as
$\alpha$ is high enough with respect to $\beta$.
Such terms (the additional ones to the classical DNLS with cubic
onsite nonlinearity and linear coupling between sites) have appeared
in physical considerations within the modeling of waveguide arrays
\cite{joha, joha3}, establishing that in the case of large penetration length or closely spaced
waveguides these terms are not negligible; however, it should be
noted that in this case additional terms of the same (cubic) order
should be included in the relevant modeling~\cite{joha,joha3}.
Nonlinear hopping terms appear also from FPU and KG chains of
anharmonic oscillators coupled with anharmonic inter-site potentials, or mixed FPU/KG chains.  The generalized DNLS system of \cite{claude} involving, among
others, the nonlinear hopping terms considered therein has been derived as a
perturbation of the integrable Ablowitz-Ladik system, by the rotating wave approximation on the FPU  chain. A similar DNLS system has been derived in \cite{joha4}, approximating the slow dynamics of the fundamental harmonic in the Fourier series expansion of discrete small amplitude modulational waves.  Relation of such DNLS systems as models for the energy transport in helical proteins has been discussed in \cite{kundu}.
However, it is worth remarking that additional terms  should also be taken into account therein, as well.
Furthermore, such terms have been studied in their own right mathematically
while considering the properties of potential traveling waves within a
generalized class of DNLS models in \cite{peli}.

\begin{figure}
\begin{center}
    \begin{tabular}{cc}
    \includegraphics[scale=0.4]{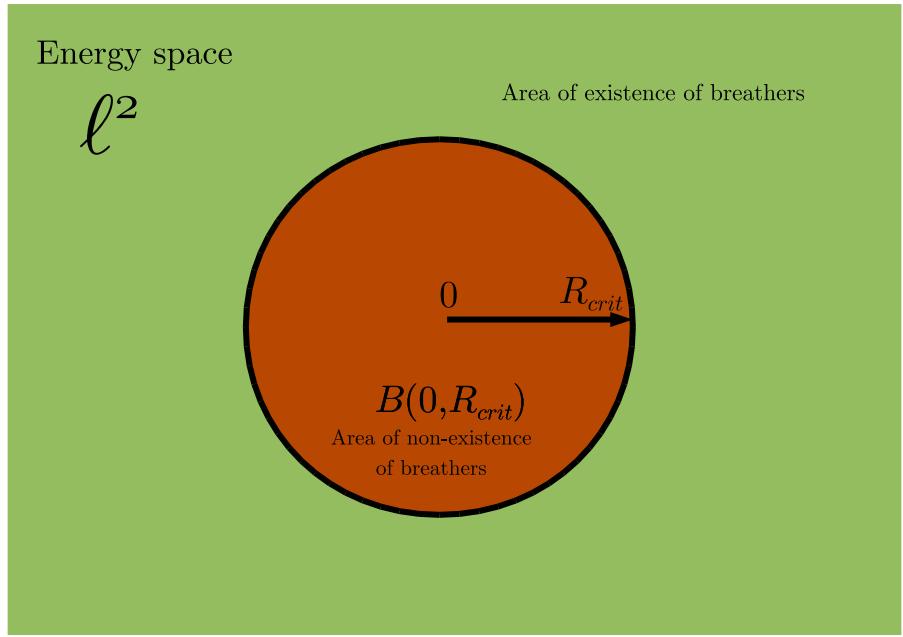}
    \end{tabular}
\caption{Simple geometric interpretation of the energy lower bounds obtained by the fixed point argument:  Breathers do not exist in the darker (red) area, the closed ball $B(0,R_{\mathrm{crit}})$ of $\ell^2$, centered at $0$ and of radius $R_{\mathrm{crit}}$. The lighter (green) area represents the area of the energy space where breather solutions exist. Although the non-existence result does not depend on the dimension and the lattice parameters, the radius $R_{\mathrm{crit}}$ of the closed ball $B(0,R_{\mathrm{crit}})$ of non-existence, quantitatively is a function of the lattice parameters $\alpha,\beta,\sigma$, the frequency $\Omega$ and the dimension of the lattice $N$. Note that $R_{\mathrm{crit}}$ is not sharp with respect to non-existence. This is suggested from the case 
%of the set of parameters 
for which the excitation threshold $R_{\mathrm{thresh}}$ is present. In this 
case it is possible that  $R_{\mathrm{thresh}}>R_{\mathrm{crit}}$, and the dark 
(red) area is enlarged.}
\label{figi}
\end{center}
\end{figure}

In this work, our  main scope is to derive lower bounds for the energy of discrete breathers for the DNLS system (\ref{hopDNLS}) and discuss their relevance as thresholds for their existence.  In this point of view, (\ref{hopDNLS}) seems to be of particular interest due to the interplay and the expected competition of the nonlinear hopping and the generalized power nonlinearities.
%This
%is true both in the case where the nonlinear hopping term is of the same order
%(in the cubic case $\sigma=1$) and in that where it is of different order (when the nonlinearity exponent $\sigma$ is increased).
Extending the arguments based on variational methods
\cite{JCN,JCN2,JFNAA010} and the fixed point approach of \cite{K1}
to establish the existence of solutions (\ref{TP}), we  show the
existence of lower bounds on the power of breathers on either finite
or infinite lattices. The bounds depend explicitly on the dimension,
and the nonlinear lattice parameters, as well as on the frequency of
the solution. They have a simple geometric  interpretation
visualized in Figure \ref{figi}, elucidated in particular by the
fixed-point approach: The energy bounds can be interpreted as the
radius $R_{\mathrm{crit}}$ of the closed ball centered at $0$ in the
energy space $\ell^2$, denoted by  $B(0,R_{\mathrm{crit}})$.
Breathers do not exist in the closed ball $B(0,R_{\mathrm{crit}})$,
and a non-trivial (e.g. non-zero) breather solution being in
$\ell^2\setminus B(0,R_{\mathrm{crit}})$ should have energy
$\mathcal{P}>R_{\mathrm{crit}}^2$. The result is of physical
significance related to energy thresholds (where by ``energy'' here
we mean power, or squared $\ell^2$ norm) for the formation of
breather solutions. In particular, it indicates that for a given set
of parameters, no periodic localized solution can have power less
than the prescribed estimates.

It should be remarked that this result is of different nature if compared with the excitation threshold phenomenon of \cite{FlachMac,Wein99} for  discrete breather families, possessing a positive lower bound on their energy when the lattice dimension
$N$  is greater than or equal to some critical dimension.  In the context of DNLS systems with power nonlinearity, the restriction for the appearance of the excitation threshold is interpreted in terms of the nonlinearity exponent as $\sigma\geq\frac{2}{N}$, \cite{Wein99}. In this point of view, $\sigma$ can be considered as critical when $\sigma=\frac{2}{N}$ and supercritical (subcritical) when $\sigma>\frac{2}{N}$ ($\sigma<\frac{2}{N}$), and the excitation threshold exists in the case $\sigma\geq\frac{2}{N}$. It is crucial to remark that the set of parameters for which the excitation threshold $\mathcal{R}_{\mathrm{thresh}}$ is apparent suggests that the energy bounds $R_{\mathrm{crit}}$ are not sharp as thresholds for existence/nonexistence. In particular, when $R_{\mathrm{crit}}$ is the value derived by the fixed-point approach, it is observed that $R_{\mathrm{thresh}}>R_{\mathrm{crit}}$, \cite{JCN, JFNAA010}.
For further discussions on the excitation threshold for FPU and Klein-Gordon lattices we refer the interested reader to \cite{Kastner2004}.

Section \upshape{II} is devoted to the derivation of the estimates by variational and energy methods employed in the case of finite lattices, and Section \upshape{III} is devoted to the fixed point approach in infinite lattices. While the methods are applicable for both subcritical and supercritical nonlinearities, in the latter case we investigate their interplay with the optimal constant of the discrete interpolation inequality of \cite{Wein99} and its analytical estimation proposed in \cite{JFN2009} (Section \upshape{IIIB}).
In Section \upshape{IV} we perform numerical simulations testing the lower bounds as thresholds for non-existence of breathers with respect to the variation of the lattice parameters, while section  \upshape{V} briefly summarizes
our conclusions. The previous studies proved the validity of these bounds as energy thresholds for the existence of breather solutions and justified that there are elements of breather families (parametrized by the lattice parameters) which tend to saturate the theoretical bounds in the case of large and small nonlinearity exponents.  Aiming to improve this prediction for extended parameter regimes, we consider a refinement of the lower bounds, on account of the finite localization length of the true breather solutions and the expectation that the main contribution to the power comes from the central and adjacent sites, being the most excited. To incorporate this claim in the numerical simulations, we perform a cut-off procedure which considers the part of the system for the oscillators occupying a unit length around the central site plus the adjacent to this unit length as well. This cut-off improves the capture of the contribution of the linear part of the system to the power, manifested in the bounds by the first eigenvalue of the linear operator. The first eigenvalue estimates the contribution of the linear part from below. Contrary to the estimation of the linear part in the real length, its unit length approximation is not negligible since the linear mode over the latter is strongly localized. This is reflected in the numerical simulations performed for the case of the cubic nonlinearity.
These simulations reveal that in the weak coupling regime the bounds are getting closer to the numerical power, and in some cases provide its accurate prediction.  This quantitative response is observed in particular versus the nonlinear hopping parameter $\alpha$. The good behavior of the estimates  indicates that the approach presented can be promising in a study of DNLS systems encountered in the aforementioned applications, involving the full expansion of nonlinear hopping terms being however of the same order.

We conclude the introductory section, by mentioning that although our results concern both the cases of finite and infinite lattices the term "breather" has been used for the standing wave solutions (\ref{TP}) in the finite case, only for the sake of brevity.  The important issue of the localization properties of the solutions in the transition from the finite to the infinite lattice is not 
addressed in the present work.  We refer to \cite{PP} for a detailed discussion on the spatial decay and stability properties of the solutions when the lattice size is varied for small-amplitudes (i.e., near the continuum limit), 
as well as, for relative localization estimates. For the convergence of 
solutions, defined by constrained variational problems in 
finite lattices to unimodal and even profile breather solutions 
(centered on a site or between two lattice sites)
in infinite lattices, we refer the interested reader to \cite{MH}.
%%%%%%%%%%%%%%%%%%%%%%%%%%%%%%%%%%%%%%%%%%%%%%%%%%%%%%%%%%%%%%%%%%%%%%%%%%%%%%%%%%
%%%%%%%%%%%%%%%%%%%%%%%%%%%%%%%%%%%%%%%%%%%%%%%%%%%%%%%%%%%%%%%%%%%%%%%%%%%%%%%5555
%%%%%%%%%%%%%%%%%%%%%%%%%%%%%%%%%%%%%%%%%%%%%%%%%%%%%%%%%%%%%%%%%%%%%%555
%%%%%%%%%%%%%%%%%%%%%%%%%%%%%%%%%%%%%%%%%%%%%%%%%%%%%%%%%%%%%%%%%%%%%%%%%
\paragraph{Preliminaries.} For convenience, we  recall from \cite{JCN,JCN2} some preliminary information on various norms and quantities, that will be thoroughly used in what follows.

The finite dimensional problem  is
formulated in the finite dimensional subspaces of the sequence spaces
$\ell^p$, $1\leq p\leq\infty$,
\begin{eqnarray}
\label{subs}
\ell^p(\mathbb{Z}^N_K)=\left\{\phi\in\ell^p\;:\;\phi_n=0\;\;\mbox{for}
  \;\;|||n|||>K\right\},
\end{eqnarray}
where $|||n|||=\max_{1\leq i\leq N}|n_i|$ for
$n=(n_1,n_2,\ldots,n_N)\in\mathbb{Z}^N$. Note that in the case of the infinite lattice $\mathbb{Z}^N$
\begin{eqnarray}
\label{lastWeinaa}
||\phi||_{q}&\leq& ||\phi||_{p},\;\;1\leq p\leq q\leq\infty\\
\label{lastWeinab}
0\leq\epsilon(-\Delta_d\phi,\phi)_{2}&\leq&
4\epsilon N \sum_{n\in\mathbb{Z}^N}|\phi_n|^2.
\end{eqnarray}
For the finite dimensional case
we have that $\ell^p(\mathbb{Z}^N_K)\equiv \mathbb{C}^{(2K+1)^N}$, endowed with the norm
\begin{eqnarray*}
||\phi||_{p}=\left(\sum_{|||n|||\leq K}|\phi_n|^p\right)^{\frac{1}{p}},
\end{eqnarray*}
and that the well known equivalence of norms,
\begin{eqnarray}
\label{fnorms}
||\phi||_{q}\leq ||\phi||_{p}\leq(2K+1)^{\frac{N(q-p)}{qp}}||\phi||_{q},\;\;1\leq p\leq q<\infty,
\end{eqnarray}
holds.

At this point let us remark on some basic facts on the eigenvalues of the discrete Dirichlet Laplacian, since they will naturally appear in the estimates that will be derived in what follows and have an important role in the numerical simulations. For the $1D$-lattice of $K+2$ oscillators, $n=0,\ldots, K+1$, let us consider the discrete eigenvalue problem for $\phi_n\in\mathbb{R}$,
\begin{eqnarray}
\label{DLap}
-\epsilon\Delta_d\phi_n=\mu \phi_n,\;\;n=1,\ldots K,
\end{eqnarray}
with Dirichlet boundary conditions, $\phi_0=\phi_{K+1}=0$. Starting from the standard case
\begin{eqnarray}
\label{case1}
\epsilon=\frac{1}{h^2}\;\;\mbox{where}\;\;h=\frac{L}{K+1},
\end{eqnarray}
where $L$ denotes the length of the chain, the eigenvalues are
\begin{eqnarray*}
\mu_n (h)=\frac{4}{h^2}\sin^2\left(\frac{n\pi h}{2L}\right)=\frac{(K+1)^2}{L^2}\sin^2\left(\frac{n\pi}{2(K+1)}\right),\;\;n=1,\ldots,K.
\end{eqnarray*}
Thus, in the case (\ref{case1}) the principal eigenvalue is
\begin{eqnarray}
\label{case1.1}
\mu_1(h)=\frac{4}{h^2}\sin^2\left(\frac{\pi h}{2L}\right)=\frac{(K+1)^2}{L^2}\sin^2\left(\frac{\pi}{2(K+1)}\right).
\end{eqnarray}
The  discrete system is modeled when $h=O(1)$, and in the limits $h\rightarrow 0$ and $h\rightarrow\infty$ we have
\begin{eqnarray}
\label{case1.2A}
&&\lim_{h\rightarrow 0}\mu_1(h) =\lambda_1=\frac{\pi^2}{L^2},\;\;\mbox{(continuous limit)},\\
\label{case1.3A}
&&\lim_{h\rightarrow \infty}\mu_1(h)=0,\;\; \mbox{(anticontinuous limit)}.
\end{eqnarray}
In the particular case of $L=1$ we have
\begin{eqnarray}
\label{case1.2}
&&\lim_{h\rightarrow 0}\mu_1(h) =\lambda_1=\pi^2,\\
\label{case1.3}
&&4\leq \mu_1 (h)\leq \pi^2,\;\;\mbox{for}\;\;0<h\leq 1.
\end{eqnarray}

In a general discrete case the parameter $\epsilon>0$ can be either related or not related with the lattice spacing $h$. As an example for the former, we may fix the linear coupling constant $\epsilon>0$, varying the number of oscillators, equidistanced with lattice spacing $h=\frac{L}{K+1}$. We have
\begin{eqnarray}
\label{case2.2}
\mu_1(h)=4\epsilon\sin^2\left(\frac{\pi h}{2L}\right)=4\epsilon\sin^2\left(\frac{\pi}{2(K+1)}\right),&&\lim_{h\rightarrow 0}\mu_1(h) =0,\;\;\mbox{($h\rightarrow 0$ when $K\rightarrow\infty$)}\\
\label{case2.3}
&&0\leq \mu_1 (h)\leq 4\epsilon.
\end{eqnarray}
Increasing $K$, (\ref{case2.2})-(\ref{case2.3}) can be considered as a particular approximation {\em of an infinite lattice}.
Note that in the case of the infinite lattice $\mathbb{Z}^N$, for the discrete Laplacian with $\epsilon=1$, we have that $\sigma(-\Delta_d)\subseteq [0,4N]$.

Relations (\ref{case1.1}) and (\ref{case1.2A}), (\ref{case1.3A}) are valid for a general coupling (depending or not depending on the lattice spacing) behaving as $\epsilon\sim\frac{1}{h^2}$ with $\epsilon$ sufficiently large. Similar observations are valid in the case of the N-dimensional discrete Laplacian.

Finally, we recall that
the variational characterization of the eigenvalues of the discrete Laplacian in the finite dimensional subspaces $\ell^2(\mathbb{Z}^N_K)$, showing that
$\mu_1>0$, can be characterized as
\begin{eqnarray}
\label{eigchar}
\mu_1=\inf_{
\begin{array}{c}
\phi \in \ell^2(\mathbb{Z}^N_K) \\
\phi \neq 0
\end{array}}\frac{(-\epsilon\Delta_d\phi,\phi)_{2}}{\sum_{|||n|||\leq K}|\phi_n|^2}.
\end{eqnarray}
Then, (\ref{eigchar}) implies the inequality
\begin{eqnarray}
\label{crucequiv}
\mu_1\sum_{|||n|||\leq K}|\phi_n|^2\leq
\epsilon(-\Delta_d\phi,\phi)_{2}\leq 4\epsilon N \sum_{|||n|||\leq K}|\phi_n|^2.
\end{eqnarray}

\section{Finite dimensional lattices}
\setcounter{equation}{0}
This section is devoted to the DNLS equation with nonlinear hopping terms $\alpha,\beta\neq 0$, supplemented  with Dirichlet boundary conditions
\begin{eqnarray}
\label{hopDNLSDC}
\mathrm{i}\dot{\psi}_n+\epsilon(\Delta_d\psi)_n&
+& \alpha\psi_n\sum_{j=1}^N(\mathcal{T}_{j}\psi)_{n\in\mathbb{Z}^N}
+\beta|\psi_n|^{2\sigma}\psi_n=0,\\
\label{BC}
\psi_n&=&0,\;||n||>K.
\end{eqnarray}
We will employ a constrained variational approach on the nonlinear energy functional involving the nonlinear hopping term. Noticing that the existence result can be established by minimization of the Hamiltonian or by application min-max methods (e.g mountain pass type theorems), the usage of alternative functionals may reveal interesting conditions on the nonlinearity parameters. An example is given in \cite[Section 2.2 \& 2.3, pg. 9--18]{JFNAA010}, where the minimization of a linear energy functional under a nonlinear constraint verified conditions for the co-existence of breather profiles. For instance, this alternative approach for (\ref{hopDNLSDC}) will show the existence of a regime for the hopping parameter $\alpha$ where an upper bound for the power is valid (see Remark  \ref{remfocA}).

Note that the case of Dirichlet boundary conditions is of interest in particular for numerical simulations; since the infinite lattice cannot be modeled numerically, numerical investigations should consider finite lattices with Dirichlet or periodic boundary conditions. The latter should be imposed for moving breathers colliding with the boundary. We expect that the variational approach can be applied in the case of periodic boundary conditions, but the details have to be checked.

We shall consider first the focusing case for the parameters $\alpha,\beta>0$ and we shall briefly comment on the defocusing one $\alpha,\beta<0$ which can be treated similarly.
%%%%%%%%%%%%%%%%%%%%%%%%%%%%%%%%%%%%%%%%%%%%%%%%%%%%%%%%%%%%%%%%%55
\subsection{The focusing case $\alpha,\beta>0$-Solutions $\psi_n(t)=e^{\mathrm{i}\Omega
  t}\phi_n,\;\Omega>0$}
Substitution of the solution (\ref{TP}) into
(\ref{hopDNLS}) shows that $\phi_n$ satisfies the system
of algebraic equations
\begin{eqnarray}
\label{SW1}
-\epsilon(\Delta_d\phi)_n+\Omega\phi_n&-&\alpha\phi_n\sum_{j=1}^N(\mathcal{T}_{j}\phi)_{n\in\mathbb{Z}^N}-\beta|\phi_n|^{2\sigma}\phi_n =0,
\;\;\Omega\in\mathbb{R},\;\;||n||\leq K,\\
\label{SW2}
\phi_n&=&0,\;||n||>K.
\end{eqnarray}
Let us note that in the anticontinuous limit $\epsilon=0$, the corresponding energy equation reads as
\begin{eqnarray*}
\Omega\sum_{||n||\leq K}|\phi_n|^2=\alpha\sum_{||n||\leq K}|\phi_n|^2\sum_{j=1}^N(\mathcal{T}_{j}\phi)_{n\in\mathbb{Z}^N}+\beta\sum_{||n||\leq K}|\phi_n|^{2\sigma+2},\;\;\alpha,\beta>0.
\end{eqnarray*}
Its positive right-hand side, implies directly that in the limit $\epsilon=0$, the focusing case supports only solutions with $\Omega>0$.

For $\epsilon>0$ we will also restrict our considerations to 
the case of solutions with $\Omega>0$.  We recall two auxiliary lemmas regarding the differentiability of the nonlinear terms if viewed as nonlinear functionals, which can be proved as in
\cite[Lemma 2.3, pg. 121]{K1}.
\begin{lemma}
\label{FDa}
Let $\phi\in\ell^{2}$. Then the functional
$$
\mathcal{V}(\phi)=\sum_{j=1}^N\sum_{n_j=-\infty}^{+\infty}|\phi_{(n_1,n_2,\ldots,n_{j},n_{j+1},\ldots,n_{N})}|^2|\phi_{(n_1,n_2,\ldots,n_{j}+1,n_{j+1},\ldots,n_{N})}|^2,
$$
is a $\mathrm{C}^{1}(\ell^{2},\mathbb{R})$ functional and for all $\psi\in\ell^{2}$,
\begin{eqnarray}
\label{gatdeva}
<\mathcal{V}'(\phi),\psi>=&&2\mathrm{Re}\sum_{j=1}^N\sum_{n_j=-\infty}^{+\infty}|\phi_{(n_1,n_2,\ldots,n_{j}+1,n_{j+1},\ldots,n_{N})}|^2\phi_{(n_1,n_2,\ldots,n_{j},n_{j+1},\ldots,n_{N})}\overline{\psi}_{(n_1,n_2,\ldots,n_{j},n_{j+1},\ldots,n_{N})}\nonumber\\
&&+2\mathrm{Re}\sum_{j=1}^N\sum_{n_j=-\infty}^{+\infty}|\phi_{(n_1,n_2,\ldots,n_{j}-1,n_{j+1},\ldots,n_{N})}|^2\phi_{(n_1,n_2,\ldots,n_{j},n_{j+1},\ldots,n_{N})}\overline{\psi}_{(n_1,n_2,\ldots,n_{j},n_{j+1},\ldots,n_{N})}.
\end{eqnarray}
\end{lemma}
%%%%%%%%%%%%%%%%%%%%%%%%%%%%%%%%%%%%%%%%%%%%%%%%%%%%%%%%%%%%%%%%%%%%%%%%%5
\begin{lemma}
\label{FDb}
Let $\phi\in\ell^{2}$. Then the functional
$$
\mathcal{L}(\phi)=\sum_{n\in\mathbb{Z}^N}|\phi_n|^{2\sigma+2}
$$
is a $\mathrm{C}^{1}(\ell^{2},\mathbb{R})$ functional and
\begin{eqnarray}
\label{gatdevb}
<\mathcal{L}'(\phi),\psi>=2(\sigma+1)\mathrm{Re}\sum_{n\in\mathbb{Z}^N}|\phi_{n}|^{2\sigma}\phi_n\overline{\psi}_n.
\end{eqnarray}
\end{lemma}
Both Lemmas \ref{FDa} and \ref{FDb} remain valid in the case of the finite lattice (space $\ell^2(\mathbb{Z}^N_K)$).

The first result on the existence of time-periodic solutions (\ref{TP}) of (\ref{hopDNLSDC}), is via a constrained minimization problem for the functional
\begin{eqnarray}
\label{enegfun1}
\mathcal{E}[\phi]:=\epsilon(-\Delta_d\phi,\phi)_2
+\Omega\sum_{n\in\mathbb{Z}^N}|\phi_n|^2-\alpha\mathcal{V}(\phi),\;\;\Omega>0,\;\;\alpha>0.
\end{eqnarray}
\setcounter{theorem}{2}
\begin{theorem}
\label{posFreq}
A. Consider the variational problem on $\ell^2(\mathbb{Z}^N_K)$
\begin{eqnarray}
\label{infsigmaE}
\inf\left\{\mathcal{E}[\phi]\;:\;\frac{1}{\sigma+1}\mathcal{L}[\phi]
=M\right\},
\end{eqnarray}
for some $\Omega>0$.
Then, there exists a minimizer $\hat{\phi}\in\ell^2(\mathbb{Z}^N_K)$ for the
variational problem (\ref{infsigmaE}) and $\beta(M)>0$, both
satisfying the Euler-Lagrange equation (\ref{SW1})-(\ref{SW2}) and
$\sum_{n\in\mathbb{Z}^N_K}|\hat{\phi}_n|^{2\sigma+2}=M(\sigma +1)$.
\newline
B. Assume that the
power of a solution of the problem (\ref{SW1})-(\ref{SW2}) is $\mathcal{P}[\hat{\phi}]=R^2$. Then the
power satisfies the lower bound
\begin{eqnarray}
\label{UBH}
R_{*,f}^2<R^2=\mathcal{P}[\hat{\phi}],
\end{eqnarray}
where $R_{*,f}$ denotes the unique positive root of the algebraic equation
\begin{eqnarray}
\label{eqR}
\beta\chi^{2\sigma}+2\alpha N\chi^2-(\mu_1+\Omega)=0.
\end{eqnarray}
C. We assume that
\begin{eqnarray}
 \label{maxsigmaA}
\sigma>1
\end{eqnarray}
Then a breather solution of (\ref{hopDNLS}) satisfies the lower bound
\begin{eqnarray}
\label{explb}
 \left[\frac{1}{2\beta}\left(\Omega+\mu_1
-\frac{(2\alpha N)^{\frac{\sigma}{\sigma-1}}}{(\beta\sigma)^{\frac{1}{\sigma-1}}}\frac{\sigma-1}{\sigma}\right)\right]^{\frac{1}{\sigma}}<R^2,
\end{eqnarray}
in either one of the cases\newline
(i) (lattice spacing condition) For all $\Omega>0$ if
\begin{eqnarray}
 \label{meshA}
\epsilon>\frac{(2\alpha N)^{\frac{\sigma}{\sigma-1}}}{(\beta\sigma)^{\frac{1}{\sigma-1}}}\frac{\sigma-1}{\lambda_1\sigma}.
\end{eqnarray}
(ii) (frequency condition) For all $\epsilon>0$ if
\begin{eqnarray}
 \label{freqmeshA}
\Omega>\frac{(2\alpha N)^{\frac{\sigma}{\sigma-1}}}{(\beta\sigma)^{\frac{1}{\sigma-1}}}\frac{\sigma-1}{\sigma}.
\end{eqnarray}
\end{theorem}
{\bf Proof:} A. We consider the set
\begin{eqnarray*}
B_{\sigma}=\left\{\phi\in\ell^2(\mathbb{Z}^N_K)\;:\;\frac{1}{\sigma+1}\mathcal{L}[\phi]=M\right\}.
\end{eqnarray*}
From Lemma \ref{gatdeva}, we may easily infer that  $\mathcal{E}:B_{\sigma}\rightarrow\mathbb{R}$ is a $C^1$-functional. Moreover, by using inequality (\ref{fnorms}), we deduce that
\begin{eqnarray*}
\mathcal{E}[\phi]&\geq& -\alpha\mathcal{V}[\phi]\\
&\geq&-\alpha N\sum_{n\in\mathbb{Z}^N}|\phi_n|^2||\phi||_2^2
\geq-\alpha N||\phi||_2^4\\
&\geq& -\alpha N(2K+1)^{\frac{2N\sigma}{\sigma+1}}(\mathcal{L}[\phi])^{\frac{2}{\sigma+1}}\\
&=&-\alpha N(2K+1)^{\frac{2N\sigma}{\sigma+1}}(M(\sigma+1))^{\frac{2}{\sigma+1}}.
\end{eqnarray*}
Therefore, the functional $\mathcal{E}:B_{\sigma}\rightarrow\mathbb{R}$ is bounded from below. By the definition of the set $B_{\sigma}$ and the fact that we are restricted to the finite dimensional space $\ell^2(\mathbb{Z}^N_K)$, it immediately follows  that any minimizing sequence associated with the variational problem (\ref{infsigmaE}) is precompact. Hence, by the Weierstra\ss\ minimization theorem \cite[Proposition 8, pg. 37]{AppZei}, any minimizing sequence has a subsequence converging to a minimizer and $\mathcal{E}$ attains its infimum at a point $\hat{\phi}$ in $B_{\sigma}$. To derive the variational equation (\ref{SW1}), we consider first the $C^1$-functional (due to Lemma \ref{FDb})
\begin{eqnarray*}
\mathcal{L}_M[\phi]=\frac{1}{\sigma+1}\mathcal{L}[\phi]-M,
\end{eqnarray*}
and we observe that for any $\phi\in B_{\sigma}$
\begin{eqnarray*}
\left<\mathcal{L}'_M[\phi],\phi\right>=2\mathcal{L}[\phi]=2M.
\end{eqnarray*}
Thus, the regular value Theorem (\cite[Section 2.9]{CJ}, \cite[Appendix
A,pg. 556]{speight}) implies that the set $B_{\sigma}=\mathcal{L}_M^{-1}(0)$
is a $C^{1}$-submanifold of $\ell^2(\mathbb{Z}^N_K)$. Application of  the
Lagrange multiplier rule, implies the existence of a parameter
$\beta=\beta(M)\in\mathbb{R}$, such that
\begin{eqnarray}
\label{ln6}
\left<\mathcal{E}'[\hat{\phi}]-\beta\mathcal{L}_M'[\hat{\phi}],
\psi\right>&=&
2\epsilon(-\Delta_d\hat{\phi},\psi)_{2}
+2\Omega\mathrm{Re}\sum_{n\in\mathbb{Z}^N}\hat{\phi}_n\overline{\psi}_n\nonumber\\
&-&2\alpha\mathrm{Re}\sum_{j=1}^N\sum_{n_j=-\infty}^{+\infty}|\hat{\phi}_{(n_1,n_2,\ldots,n_{j}+1,n_{j+1},\ldots,n_{N})}|^2\hat{\phi}_{(n_1,n_2,\ldots,n_{j},n_{j+1},\ldots,n_{N})}\overline{\psi}_{(n_1,n_2,\ldots,n_{j},n_{j+1},\ldots,n_{N})}\nonumber\\
&-&2\alpha\mathrm{Re}\sum_{j=1}^N\sum_{n_j=-\infty}^{+\infty}|\hat{\phi}_{(n_1,n_2,\ldots,n_{j}-1,n_{j+1},\ldots,n_{N})}|^2\hat{\phi}_{(n_1,n_2,\ldots,n_{j},n_{j+1},\ldots,n_{N})}\overline{\psi}_{(n_1,n_2,\ldots,n_{j},n_{j+1},\ldots,n_{N})}\nonumber\\
&-&2\beta\mathrm{Re}\sum_{n\in\mathbb{Z}^N}
|\hat{\phi}_n|^{2\sigma}\hat{\phi}_n\overline{\psi}_n=0,\;\;\mbox{for all}\;\;
\psi\in\ell^2(\mathbb{Z}^N_K).
\end{eqnarray}
Setting $\psi=\hat{\phi}$ in (\ref{ln6}), we find that
\begin{eqnarray}
\label{ln7}
\mathcal{F}[\hat{\phi}]&:=&\epsilon (-\Delta_d\hat{\phi},\hat{\phi})_2+\Omega\sum_{n\in\mathbb{Z}^N}|\hat{\phi}_n|^2\nonumber\\
&-&2\alpha\mathrm{Re}\sum_{j=1}^N\sum_{n_j=-\infty}^{+\infty}|\hat{\phi}_{(n_1,n_2,\ldots,n_{j}+1,n_{j+1},\ldots,n_{N})}|^2|\hat{\phi}_{(n_1,n_2,\ldots,n_{j},n_{j+1},\ldots,n_{N})}|^2\nonumber\\
&-&2\alpha\mathrm{Re}\sum_{j=1}^N\sum_{n_j=-\infty}^{+\infty}|\hat{\phi}_{(n_1,n_2,\ldots,n_{j}-1,n_{j+1},\ldots,n_{N})}|^2|\hat{\phi}_{(n_1,n_2,\ldots,n_{j},n_{j+1},\ldots,n_{N})}|^2\nonumber\\
&=&\beta\sum_{n\in\mathbb{Z}^N}|\hat{\phi}_{n}|^{2\sigma+2}.
\end{eqnarray}
By virtue of (\ref{crucequiv}), we deduce that the following estimate
\begin{eqnarray}
\label{ln8}
\mathcal{F}[\phi]&\geq&\mu_1||\hat{\phi}||_2^2+\Omega||\hat{\phi}||_2^2-2\alpha N\sum_{n\in\mathbb{Z}^N}||\hat{\phi}||^2_2|\hat{\phi}_n|^2\nonumber\\
&\geq&\mu_1||\hat{\phi}||_2^2+\Omega||\hat{\phi}||_2^2-2\alpha N||\hat{\phi}||_2^4,
\end{eqnarray}
holds. Let us assume that $\mathcal{P}[\hat{\phi}]=||\hat{\phi}||_2^2=R^2$. Then from (\ref{ln8}), we obtain that
\begin{eqnarray*}
\mathcal{F}[\phi]\geq R^2(\mu_1+\Omega-2\alpha N R^2).
\end{eqnarray*}
%%%%%%%%%%%%%%%%%%%%%%%%%%%%%%%%%%%%%%%%%%%%%%%%%%%%%%%%%%%%%%%%%%%%%%%%%%%%%%%%%
%%%%%%%%%%%%%%%%%%%%%%%%%%%%%%%%%%%%%%%%%%%%%%%%%%%%%%%%%%%%%%%%%%%%%%%%%%%%%%%5555
Therefore, assuming that
\begin{eqnarray}
\label{UB}
R^2<\frac{\mu_1+\Omega}{2\alpha N},
 \end{eqnarray}
or assuming in terms of $\alpha$ that
\begin{eqnarray}
\label{UBA}
0<\alpha<\frac{\mu_1+\Omega}{2NR^2},
 \end{eqnarray}
we deduce that $\mathcal{F}[\hat{\phi}]>0$.
Since $\hat{\phi}\in B_{\sigma}$ cannot be identically zero and
$\mathcal{F}[\hat{\phi}]>0$, it follows from (\ref{ln7}) that
$\beta>0$. Summarizing, we have proved that for given $\Omega>0$, there exists a minimizer $\hat{\phi}$ and a Lagrange multiplier $\beta>0$ solving the variational equation (\ref{ln6}). Clearly a solution of the variational equation (\ref{ln6}) is a solution of the Euler-Lagrange equation (\ref{SW1})-(\ref{SW2}).
\newline
B. It is necessary to verify first that any solution  $\hat{\phi}$ of (\ref{SW1})-(\ref{SW2}) is a solution of the minimization problem (\ref{infsigmaE}). Indeed,  if $\hat{\phi}$ is a solution of (\ref{SW1})-(\ref{SW2}), multiplying (\ref{SW1}) by $\hat{\phi}$ in the $\ell^2(\mathbb{Z}^N_K)$ and using the Dirichlet boundary conditions we infer that $\hat{\phi}$ satisfies  equation (\ref{ln7}), written as
\begin{eqnarray}
\label{ln7A}
\mathcal{F}[\hat{\phi}]=\beta\mathcal{L}[\hat{\phi}].
\end{eqnarray}
Then, due to Lemmas \ref{FDa} and \ref{FDb}, $\hat{\phi}$ solves also the equation
\begin{eqnarray*}
\label{var2}
\left<\mathcal{F}'[\hat{\phi}],\psi\right>=\beta\left<\mathcal{L}'[\hat{\phi}],\psi\right>,\;\;\mbox{for all}\;\;\psi\in\ell^2(\mathbb{Z}^N_K).
\end{eqnarray*}
Comparing (\ref{ln6}) with (\ref{ln7}) it can be easily seen that the equation above is equivalent to
\begin{eqnarray}
\label{var2a}
\left<\mathcal{E}'[\hat{\phi}],\psi\right>=\beta\left<\mathcal{L}'[\hat{\phi}],\psi\right>,\;\;\mbox{for all}\;\;\psi\in\ell^2(\mathbb{Z}^N_K),
\end{eqnarray}
thus, $\hat{\phi}$ is a minimizer of the minimization problem (\ref{infsigmaE}). The converse follows immediately by (\ref{var2a}) and the fact that in the discrete setting a ``weak solution'' of (\ref{var2a}) coincides with a solution of  (\ref{SW1})-(\ref{SW2}). Furthermore, by setting $\psi=\hat{\phi}$ in (\ref{var2a}) we recover that $\hat{\phi}$ satisfies the equation (\ref{ln7A}).

Assuming now that the power of the solution of (\ref{SW1}) is $\mathcal{P}[\hat{\phi}]=||\hat{\phi}||_2^2=R^2$, by using (\ref{fnorms}) and
(\ref{crucequiv}) we get from (\ref{ln7A}), that $R$ satisfies the inequality
\begin{eqnarray}
\label{LB}
\mu_1+\Omega\leq 2\alpha N R^2+\beta R^{2\sigma}.
\end{eqnarray}
The algebraic equation (\ref{eqR}) considered for $\chi\in [0,\infty)$,  has exactly one positive root $0<R_{*,f}$. Then, comparison of the equation (\ref{eqR}) with inequality (\ref{LB}), implies that the power $\mathcal{P}[\hat{\phi}]$ must satisfy the lower bound
(\ref{UBH}).
\newline
C. Applying Young's inequality
$$ab<\frac{\hat{\epsilon}}{p}a^p+\frac{1}{q\hat{\epsilon}^{q/p}}b^q,\;\;a,b>0\;\;\mbox{for any}\;\;\hat{\epsilon}>0, \;\;1/p+1/q=1,$$
with $p=\sigma$, $q=\frac{\sigma}{\sigma-1}$ $a=R^2$, $b=2\alpha N$ and
$\hat{\epsilon}=\beta\sigma$
we get that
\begin{eqnarray}
\label{claim4}
2\alpha R^2\leq \beta R^{2\sigma}+\frac{(2\alpha N)^{\frac{\sigma}{\sigma-1}}}{(\beta\sigma)^{\frac{1}{\sigma-1}}}
\frac{\sigma-1}{\sigma}
\end{eqnarray}
Inserting (\ref{claim4}) into (\ref{LB}) we derive the lower bound (\ref{explb}).
\ \ $\diamond$
%%%%%%%%%%%%%%%%%%%%%%%%%%%%%%%%%%%%%%%%%%%%%%%%%%%%%%%%%%%%%%%%%%%%%%%%%%%%%%%%%%%%%%%%%%%%%%%%%%%%%%%%%%%%%%%%%%%%
%%%%%%%%%%%%%%%%%%%%%%%%%%%%%%%%%%%%%%%%%%%%%%%%%%%%%%%%%%%%%%%%%%%%%%%%%%%%%%%%%%%%%%%%%%%%%%%%%%%%%%%%%%%%%%%%%%%%
%%%%%%%%%%%%%%%%%%%%%%%%%%%%%%%%%%%%%%%%%%%%%%%%%%%%%%%%%%%%%%%%%%%%%%%%%%%%%%%%%%%%%%%%%%%%%%%%%%%%%%%%%%%%%%%%%%%%%%%
%%%%%%%%%%%%%%%%%%%%%%%%%%%%%%%%%%%%%%%%%%%%%%%%%%%%%%%%%%%%%%%%%%%%%%%%%%%%%%%%%%%%%%%%%%%%%%%%%%%%%%%%%%%%%%%%%%%%%%%
\setcounter{remark}{2}
\begin{remark}
\label{remfocA}
{\em 1. (The lower bound for the cubic nonlinearity) For the case of cubic nonlinearity $\sigma=1$, inequality (\ref{LB}) implies that the power of the periodic solution $\psi_n(t)=e^{i\Omega t}\hat{\phi}_n$, $\Omega>0$ must satisfy the lower bound
\begin{eqnarray}
\label{cubicLB}
\frac{\mu_1+\Omega}{2\alpha N +\beta}<R^2=\mathcal{P}[\hat{\phi}].
\end{eqnarray}
\newline
2. (Interpretation of condition (\ref{UB})). The result of Theorem (\ref{posFreq}) establishes for arbitrary given $\Omega>0$ and $\alpha>0$, the existence of a nontrivial $\hat{\phi}\in\ell^2(\mathbb{Z}^N_K)$ and the existence of $\beta>0$ as a Lagrange mulitplier such that $\psi_n(t)=e^{i\Omega t}\hat{\phi}_n$, solves equation (\ref{hopDNLSDC}) with $\beta>0$ as a parameter for the power nonlinearity. On the account of this result, the meaning of condition (\ref{UB}) is {\em that there exists $\beta>0$  and a range of the hopping parameter $0<\alpha<\alpha^*$ for which the associated minimizer
$\hat{\phi}$ has power satisfying the upper bound}
\begin{eqnarray}
\label{possRegion}
\mathcal{P}[\hat{\phi}]=R^2<\frac{\mu_1+\Omega}{2\alpha N}.
\end{eqnarray}
Note that the existence of the range of the hopping parameter $\alpha$ stated above is also established by (\ref{UB})-see (\ref{UBA}).
\newline
3. (Case $\alpha\rightarrow 0,\beta>0$-DNLS with power nonlinearity). The proof of Theorem (\ref{posFreq}) remains valid for the case $\alpha=0$, where one has to consider the constrained minimization problem (\ref{infsigmaE}) for the functional $\mathcal{E}$, setting $\alpha=0$. Thus for the classical DNLS with power nonlinearity we recover from inequality (\ref{LB}), the lower bound
\begin{eqnarray}
\label{classDNLS}
\left[\frac{\mu_1+\Omega}{\beta}\right]^{\frac{1}{\sigma}} <R^2=\mathcal{P}[\hat{\phi}].
\end{eqnarray}
The lower bound (\ref{classDNLS}) is the same as (5.27) and (5.31) of \cite{JCN} for the DNLS with power nonlinearity.
}
\end{remark}
%%%%%%%%%%%%%%%%%%%%%%%%%%%%%%%%%%%%%%%%%%%%%%%55555
%%%%%%%%%%%%%%%%%%%%%%%%%%%%%%%%%%%%%%%%%%%%%%%%%%%
\subsection{The defocusing case $\alpha,\beta<0$-Solutions $\psi_n(t)=e^{\mathrm{-i}\Omega
  t}\phi_n,\;\Omega>0$.}
We shall briefly comment on the existence of breather solutions, for the case of negative nonlinear parameters $\alpha,\beta<0$. We set for convenience $\alpha=-\kappa,\,\beta=-\lambda$ where $\kappa,\lambda>0$.
It should be remarked that the case of negative parameters can be reduced to the case of positive ones, under the  staggering transformation. We recall that this transformation is defined as
\begin{eqnarray}
\label{stag}
\psi_n\rightarrow (-1)^{|n|}\psi_n,\;\;\; |n|=\sum_{i=1}^N n_i,
\end{eqnarray}
(see e.g. the discussion of \cite[pg. 7]{Panos3}).
The case of negative parameters, corresponds to the existence problem for solutions
\begin{eqnarray}
\label{bhopn}
\psi_n(t)=e^{\mathrm{-i}\Omega
  t}\phi_n,\;\Omega>0,
\end{eqnarray}
where $\phi_n$ satisfies the system
of algebraic equations
\begin{eqnarray}
\label{SW1a}
-\epsilon(\Delta_d\phi)_n-\Omega\phi_n&+&\kappa\phi_n\sum_{j=1}^N(\mathcal{T}_{j}\phi)_{n\in\mathbb{Z}^N}+\lambda|\phi_n|^{2\sigma}\phi_n =0,
\;\;\Omega>0,\;\;||n||\leq K,\\
\label{SW2a}
\phi_n&=&0,\;||n||>K.
\end{eqnarray}
The proof of the existence of breather solutions (\ref{bhopn}) is very similar to that of Theorem \ref{posFreq}, and we refrain from giving the details. We just note that the constrained minimization problem will  consider the $C^1$-functional
\begin{eqnarray}
\label{enegfun1a}
\mathcal{E}[\phi]:=\epsilon(-\Delta_d\phi,\phi)_2
-\Omega\sum_{n\in\mathbb{Z}^N}|\phi_n|^2+\kappa\mathcal{V}(\phi),\;\;\Omega>0,\;\;\kappa>0.
\end{eqnarray}
\setcounter{theorem}{3}
\begin{theorem}
\label{negFreq}
A. Consider the variational problem on $\ell^2(\mathbb{Z}^N_K)$
\begin{eqnarray}
\label{infsigmaE2}
\inf\left\{\mathcal{E}[\phi]\;:\;\frac{1}{\sigma+1}\mathcal{L}[\phi]
=M\right\}.
\end{eqnarray}
for some $\Omega>0$. Assume further that
\begin{eqnarray}
\label{conOmega}
\Omega>4\epsilon N.
\end{eqnarray}
Then, there exists a minimizer $\phi^*\in\ell^2(\mathbb{Z}^N_K)$ for the
variational problem (\ref{infsigmaE2}) and $\lambda(M)>0$,
satisfying both the Euler-Lagrange equation (\ref{SW1a})-(\ref{SW2a}) and
$\sum_{n\in\mathbb{Z}^N}|{\phi}_n^*|^{2\sigma+2}=M$.
\newline
B. Assume that (\ref{conOmega}) holds and that  the
power of a solution of the problem (\ref{SW1a})-(\ref{SW2a}) is $\mathcal{P}[{\phi}^*]=R^2$. Then the
power satisfies the lower bound
\begin{eqnarray}
\label{UBHa}
R_{*,d}^2<R^2=\mathcal{P}[\hat{\phi}],
\end{eqnarray}
where $R_{*,d}$ denotes the unique positive root of the  equation
\begin{eqnarray}
\label{eqRa}
\lambda\chi^{2\sigma}+2\kappa N\chi^2-(\Omega-4\epsilon N)=0.
\end{eqnarray}
C. Let $\sigma>1$ and assume that
\begin{eqnarray}
\label{explRN}
\Omega>4\epsilon N+\frac{\sigma-1}{\lambda^{\frac{1}{\sigma-1}}}\left(\frac{2\kappa N}{\sigma}\right)^{\frac{\sigma}{\sigma-1}}
\end{eqnarray}
Then the power satisfies the lower bound
\begin{eqnarray}
\label{explbN}
\left[\frac{1}{2\lambda}\left(\Omega-4\epsilon N
-\frac{(2\kappa N)^{\frac{\sigma}{\sigma-1}}}{(\lambda\sigma)^{\frac{1}{\sigma-1}}}\frac{\sigma-1}{\sigma}\right)\right]^{\frac{1}{\sigma}}<R^2=\mathcal{P}[\hat{\phi}]
\end{eqnarray}
\end{theorem}
\setcounter{remark}{4}
\begin{remark}

{\em 1. (The lower bound for the cubic nonlinearity) For the case of negative parameters $\alpha=-\kappa,\beta=-\lambda$, $\kappa,\lambda>0$ and of cubic nonlinearity $\sigma=1$, the power of the periodic solution $\psi_n(t)=e^{-i\Omega t}\hat{\phi}_n$, $\Omega>0$ must satisfy the lower bound
\begin{eqnarray}
\label{cubicLBan}
\frac{\Omega-4\epsilon N}{2\kappa N +\lambda}<R^2=\mathcal{P}[\phi^*],\;\;\Omega>4\epsilon N
\end{eqnarray}
\newline
2. (An upper bound for some range of parameters). The result of Theorem \ref{negFreq} establishes for given $\Omega>4\epsilon$ and $\alpha=-\kappa<0$, the existence of a nontrivial $\phi^*\in\ell^2(\mathbb{Z}^N_K)$ and the existence of $\beta=-\lambda<0$ such that $\psi_n(t)=e^{-i\Omega t}\phi^*_n$, solves equation (\ref{hopDNLSDC}) with $\beta>0$ as a parameter for the power nonlinearity. As in remark \ref{remfocA}-2, a similar condition to (\ref{UB}) can be derived, implying that there exists a parameter $\lambda$ and a range for the hopping parameter $\kappa$ for which the corresponding minimizer $\phi^*$ has power satisfying the upper bound
\begin{eqnarray}
\label{possRegionn}
\mathcal{P}[\phi^*]<\frac{\Omega-4\epsilon N}{2\kappa N},\;\;\Omega>4\epsilon N.
\end{eqnarray}
3. (Case $\kappa\rightarrow 0,\lambda>0$-DNLS with {\em defocusing} power nonlinearity). The proof of Theorem \ref{negFreq} remains valid for the case $\kappa=0$, where one has to consider the constrained minimization problem (\ref{infsigmaE2}) for the functional $\mathcal{E}$, setting $\kappa=0$. Thus for the classical DNLS with power nonlinearity we recover  the lower bound
\begin{eqnarray}
\label{classDNLSa}
\left[\frac{\Omega-4\epsilon N}{\lambda}\right]^{\frac{1}{\sigma}} <R^2=\mathcal{P}[\phi^*],\;\;\Omega>4\epsilon N.
\end{eqnarray}
The lower bound (\ref{classDNLS}) is exactly the same with that derived in \cite{JCN} for the one dimensional DNLS with  defocusing power nonlinearity.
\newline
4. Condition (\ref{conOmega}) is related with the extension of the phonon band for defocusing-type DNLS equations,
to the interval $[0, 4\epsilon N]$. Combining the results of Therorem \ref{posFreq} for the focusing case and of Theorem \ref{negFreq} for the defocusing one, we have that for breathers in the ansatz $\psi_n=e^{-\mathrm{i}\Omega t}\phi_n$, frequencies $\Omega\in\mathbb{R}$, must lie in the intervals $\Omega>4\epsilon N$ (defocusing case) and $\Omega<0$ (focusing case).}
\end{remark}
%%%%%%%%%%%%%%%%%%%%%%%%%%%%%%%%%%%%%%%%%%%%%%%%%%555555
%%%%%%%%%%%%%%%%%%%%%%%%%%%%%%%%%%%%%%%%%%%%%%%%%%%%%%%%
\section{Infinite $\mathbb{Z}^N$, $N\geq 1$ lattices}
\label{enot3}
\setcounter{equation}{0}
For the infinite lattice $\mathbb{Z}^N$, we will consider the problem of energy bounds for breathers of the DNLS (\ref{hopDNLS}) by a fixed-point method. The method establishes that the stationary problem (\ref{SW1}) defines a locally Lipschitz map on the phase space $\ell^2$. When the map is a contraction, gives rise only to the trivial solution. The Lipschitz constant for the contraction mapping defines the critical power above which we should expect existence of breathers. Below this critical power there is non-existence of breather solutions. The Lipschitz constant contains all the lattice parameters, including the dimension of the lattice and the frequency of the solution.

%%%%%%%%%%%%%%%%%%%%%%%%%%%%%%%%%%%%%%%%%%%%%%%%%%%%%%%%%%%%%%%%%%%%%%%%%%%%%%%%%%%%%%%%%%%%%%%%%%%%%%%%%%%%%%%%%%%%%%%%%%%%%%%%%%
\subsection{The case $\alpha,\beta>0$-Solutions $\psi_n(t)=e^{\mathrm{i}\Omega t}$, $\Omega>0$: Fixed point method}
The infinite system of algebraic equations  (\ref{SW1}) for breathers in the case of the infinite lattice
will be treated by a fixed point argument.
We recall that the linear and continuous operator
\begin{eqnarray}
\label{r2H}
-\epsilon\Delta_d+\Omega: \ell^2\rightarrow\ell^2,
\end{eqnarray}
satisfies  the assumptions  of Lax-Milgram Theorem \cite[Theorem 18.E, pg. 68]{zei85}, since
\begin{eqnarray*}
\epsilon(-\Delta_d\phi,\phi)_2+\Omega ||\phi||^2_2\geq \Omega||\phi||^2_2\;\;\mbox{for all}\;\;\phi\in\ell^2.
\end{eqnarray*}
This is the first step to verify that for given $z\in\ell^2$, the {\em auxiliary problem} defined by the linear operator equation
\begin{eqnarray}
\label{linearH}
-\epsilon\Delta_d\phi_n+\Omega\phi_n=\alpha z_n\sum_{j=1}^N(\mathcal{T}_{j}z)_{n\in\mathbb{Z}^N}+\beta|z_n|^{2\sigma}z_n,\
\end{eqnarray}
has a unique solution $\phi\in\ell^2$. The second step, according the Lax-Milgram Theorem is to justify that the right hand side of (\ref{linearH}) is in $\ell^2$ if $z\in\ell^2$.
Indeed, by using the inequality
\begin{eqnarray}
 \label{com1h}
\sum_{n\in\mathbb{Z}^N}| \phi_n|^p\leq \left(\sum_{n\in\mathbb{Z}^N}| \phi_n|^q\right)^{\frac{p}{q}},\;\;\mbox{for all}\;\; 1\leq q\leq p\leq\infty,
\end{eqnarray}
for $p=4\sigma+2$ and $q=2$, it follows that
\begin{eqnarray}
\label{lmeh}
|||z|^{2\sigma}z||^2_2\leq \sum_{n\in\mathbb{Z}^N}|z_n|^{4\sigma +2}
\leq ||z||_2^{4\sigma +2}.
\end{eqnarray}
Furthermore, for the nonlinear map $\mathcal{J}:\ell^2\rightarrow\ell^2$,
\begin{eqnarray*}
 \mathcal{J}[z_n]=z_n\sum_{j=1}^N(\mathcal{T}_{j}z)_{n\in\mathbb{Z}^N},
\end{eqnarray*}
we have
\begin{eqnarray*}
 ||\mathcal{J}[z]||_2^2\leq 2N\sup_{n\in\mathbb{Z}^N}|z_n|^2\sum_{n\in\mathbb{Z}^N}|z_n|^2\leq 2N||z||^4_2.
\end{eqnarray*}
Therefore  we are allowed to define the map
$\mathcal{A}:\ell^2\rightarrow\ell^2$, by $\mathcal{A}(z):=\phi$, where $\phi$ is a unique solution of the operator equation (\ref{linearH}). Clearly the map $\mathcal{A}$ is well defined. Let $\zeta,\xi$ be in the closed ball
$$B_R:=\{z\in\ell^2\;:||z||_{\ell^2}\leq R\},$$
and $\phi=\mathcal{A}(\zeta)$, $\psi=\mathcal{A}(\xi)$. The difference $\chi:=\phi-\psi$ satisfies the equation
\begin{eqnarray}
\label{claim2h}
-\epsilon\Delta_d\chi_n+\Omega\chi_n=\alpha \left(\mathcal{J}[\zeta_n]-\mathcal{J}[\xi_n]\right)+\beta(|\zeta_n|^{2\sigma}\zeta_n-|\xi_n|^{2\sigma}\xi_n).
\end{eqnarray}
We consider the linear and continuous operator $\mathcal{M}:\ell^2\rightarrow\ell^2$
\begin{eqnarray*}
 \mathcal{M}[z_n]=\sum_{j=1}^N z_{(n_1,n_2,\ldots,n_{j}+1,n_{j+1},\ldots,n_{N})}+z_{(n_1,n_2,\ldots,n_{j}-1,n_{j+1},\ldots,n_{N})},\;\;j=1,\ldots,N
\end{eqnarray*}
satisfying
\begin{eqnarray}
 \label{liph}
||\mathcal{M}[\phi]-\mathcal{M}[\psi]||_2\leq 2N||\phi-\psi||,\;\;\mbox{for all}\;\;\phi,\psi\in\ell^2.
\end{eqnarray}
Then, the first term of the right-hand side of (\ref{claim2h}) can be written as
\begin{eqnarray*}
 \alpha(\mathcal{J}[\xi_n]-\mathcal{J}[\zeta_n])=\alpha \mathcal{M}[|\xi_n|^2](\xi_n-\zeta_n)
+\alpha\zeta_n\left(\mathcal{M}[|\xi_n|^2]-\mathcal{M}[|\zeta_n|^2]\right).
\end{eqnarray*}
By using (\ref{liph}) and inequality (\ref{com1h}) for $p=4$ and $q=2$, we observe that
\begin{eqnarray}
 \label{liph1}
|| \mathcal{M}[|\xi|^2](\xi-\zeta)||_2^2&=&\sum_{n\in\mathbb{Z}^N}\mathcal{M}^2[|\xi_n|]|\xi_n-\zeta_n|^2\nonumber\\
&\leq& \sup_{n\in\mathbb{Z}^N}|\mathcal{M}[|\xi_n|^2]|^2\sum_{n\in\mathbb{Z}^N}|\xi_n-\zeta_n|^2\nonumber\\
&\leq& 4N^2\sum_{n\in\mathbb{Z}^N}|\xi_n|^4\sum_{n\in\mathbb{Z}^N}|\xi_n-\zeta_n|^2\nonumber\\
&\leq& 4N^2\left(\sum_{n\in\mathbb{Z}^N}|\xi_n|^2\right)^2\sum_{n\in\mathbb{Z}^N}|\xi_n-\zeta_n|^2\nonumber\\
&\leq&4N^2R^4||\xi-\zeta||_2^2.
\end{eqnarray}
Using again (\ref{liph1}) we get that
\begin{eqnarray}
 \label{liph2}
||\zeta\left(\mathcal{M}[|\xi|^2]-\mathcal{M}[|\zeta|^2]\right)||^2_2&=&\sum_{n\in\mathbb{Z}^N}|\zeta_n|^2|\mathcal{M}[|\xi_n|^2]-\mathcal{M}[|\zeta_n|^2]|^2\nonumber\\
&\leq& 4N^2\sup_{n\in\mathbb{Z}^N}|\zeta_n|^2\sum_{n\in\mathbb{Z}^N}||\xi_n|^2-|\zeta_n|^2|\nonumber\\
&\leq& 4N^2R^2\sup_{n\in\mathbb{Z}^N}(|\xi_n|+|\zeta_n|)^2\sum_{n\in\mathbb{Z}^N}|\xi_n-\zeta_n|^2\nonumber\\
&\leq& 8N^2R^4||\xi-\zeta||^2_2.
\end{eqnarray}
Hence, from (\ref{liph1}) and (\ref{liph2}), the inequality
\begin{eqnarray}
 \label{liph3}
||\mathcal{J}[\xi]-\mathcal{J}[\zeta]||_2\leq \sqrt{12}NR^2||\xi-\zeta||_2
\end{eqnarray}
readily follows. Moreover, it holds that (cf. \cite[Lemma II.2]{JFN2009})
\begin{eqnarray}
\label{liph4}
\sum_{n\in\mathbb{Z}^N}||\zeta_n|^{2\sigma}\zeta_n-|\xi_n|^{2\sigma}\xi_n|^2\leq (2\sigma +1)^2R^{4\sigma}\sum_{n\in\mathbb{Z}^N}|\zeta_n-\xi_n|^2.
\end{eqnarray}
Now, taking the scalar product of (\ref{claim2h}) with $\chi$ in $\ell^2$ and using (\ref{liph3}) and (\ref{liph4}), we have
\begin{eqnarray}
\label{cmap1a}
\epsilon(-\Delta_d\chi,\chi)_2+\Omega ||\chi||^2_2&\leq&\alpha||\chi||_2||\mathcal{J}[\xi]-\mathcal{J}[\zeta]||_2+ \beta||\chi||_2||\,|\zeta|^{2\sigma}\zeta-|\xi|^{2\sigma}\xi||_2\nonumber\\
&\leq& L(R)||\chi||_2||\zeta-\xi||_2,
\end{eqnarray}
where
\begin{eqnarray*}
 L(R)=\sqrt{12}\alpha NR^2+\beta(2\sigma+1)R^{2\sigma}.
\end{eqnarray*}
Since $(-\Delta_d\chi,\chi)_2\geq 0$, from (\ref{cmap1a}) we get the inequality
\begin{eqnarray}
\label{claim4h}
\Omega||\chi||_2^2
\leq \frac{L^2(R)}{2\Omega}||\zeta-\xi||^2_2+\frac{\Omega}{2}||\chi||_2^2.
\end{eqnarray}
From (\ref{claim4h}), we conclude that
\begin{eqnarray*}
||\chi||_2^2=||\mathcal{A}(z)-\mathcal{A}(\xi)||^2_2
\leq \frac{L^2(R)}{\Omega^2}||\zeta-\xi||^2_2,
\end{eqnarray*}
and, hence, the map  $\mathcal{A}:B_R\rightarrow B_R$ is Lipschitz continuous with the Lipschitz constant
$$M(R)=\frac{L(R)}{\Omega}.$$

The map $\mathcal{A}$ is a contraction, and hence, has a unique fixed point if
\begin{eqnarray}
\label{claim5h}
M(R)<1.
\end{eqnarray}
This unique fixed point is the trivial one, since $\mathcal{A}(0)=0$.
We consider the polynomial function
\begin{eqnarray}
\label{claim6h}
\Pi(R):=L(R)-\Omega.
\end{eqnarray}
The threshold value for the existence of nontrivial breather solutions
can be derived from condition (\ref{claim5h}), as in the proof of Theorem \ref{posFreq}B:
Denote by $R_{\mathrm{crit}}$ the positive root of the polynomial equation $\Pi(R)=0$. Then
$\Pi(R)<0$ for every $R\in(0,R_{\mathrm{crit}})$,
that is, condition (\ref{claim5h}) is satisfied if $R\in (0, R_{\mathrm{crit}})$. Therefore breathers of arbitrary energy do not exist. A breather should have
power  $R^2>R_{\mathrm{crit}}^2$. We summarize in
\begin{theorem}\label{notrih}
We assume that the parameters $\alpha, \beta,\sigma>0$
Let $R_{\mathrm{crit}}>0$ denote the unique positive root of the polynomial equation
$\Pi(R)=0$, where  $\Pi(R)$ is given by (\ref{claim6h})
Then a breather solution $\psi_n(t)=e^{\mathrm{i}\Omega t}\phi_n$, for any $\Omega>0$ of (\ref{hopDNLS})
must have power $\mathcal{P}>R_{\mathrm{crit}}^2$.
\end{theorem}

The simple geometric interpretation of Theorem \ref{notrih} is visualized in Figure \ref{figi}. Breathers do not exist in the sphere $B(0,R_{\mathrm{crit}})$ of the energy space $\ell^2$.
%%%%%%%%%%%%%%%%%%%%%%%%%%%%%%%%%%%%%%%%%%%%%%%%%%%%%%%%%%%%%%%%%%%%%%%%%%%%%%%%
\subsection{Estimates for supercritical nonlinearity exponents $\sigma\geq 2/N$.}
A different version of dimension-dependent estimates in the case of the infinite lattice can be produced by using the discrete interpolation inequality of \cite{Wein99}
\begin{eqnarray}
\label{WGN1}
\sum_{n\in\mathbb{Z}^N}|\phi_n|^{2\sigma+2}\leq
C_*\left(\sum_{n\in\mathbb{Z}^N}| \phi_n|^2\right)^{\sigma}(-\Delta_d\phi,\phi)_2,\;\;\sigma\geq\frac{2}{N}.
\end{eqnarray}
However, since (\ref{WGN1}) is valid only for $\sigma\geq N/2$, \textit{the derived estimates will refer only to this range of parameters}. We recall that the range $\sigma\geq N/2$ is related to the appearance of the excitation threshold for breathers on DNLS lattices with power law nonlinearity.

We start by multiplying (\ref{SW1}) by $\phi$ and summing over $\mathbb{Z}^N$, to get the equation
\begin{eqnarray}
 \label{dd1}
\epsilon (-\Delta_d\phi, \phi)_2+\Omega\sum_{n\in\mathbb{Z}^N}|\phi_n|^2=\alpha\sum_{n\in\mathbb{Z}^N}\sum_{j=1}^N|\phi_n|^2(\mathcal{T}_{j}\phi)_{n\in\mathbb{Z}^N}+\beta\sum_{n\in\mathbb{Z}^N}|\phi_n|^{2\sigma+2}.
\end{eqnarray}
Using (\ref{WGN1}) in order to estimate the $(-\Delta_d\phi, \phi)_2$ term of (\ref{dd1}) we have
\begin{eqnarray}
\label{dd2}
 \frac{\epsilon}{C_*}\frac{\sum_{n\in\mathbb{Z}^N}|\phi_n|^{2\sigma+2}}{\left(\sum_{n\in\mathbb{Z}^N}
|\phi_n|^2\right)^{\sigma}}+\Omega\sum_{n\in\mathbb{Z}^N}|\phi_n|^2&\leq&\alpha\sum_{n\in\mathbb{Z}^N}\sum_{j=1}^N|\phi_n|^2(\mathcal{T}_{j}\phi)_{n\in\mathbb{Z}^N}+\beta\sum_{n\in\mathbb{Z}^N}|\phi_n|^{2\sigma+2}\nonumber\\
&\leq&2\alpha N\left(\sum_{n\in\mathbb{Z}^N}|\phi_n|^2\right)^2+\beta\sum_{n\in\mathbb{Z}^N}|\phi_n|^{2\sigma+2}.
\end{eqnarray}
The inequality (\ref{dd2}) can be rewritten as
\begin{eqnarray}
 \label{dd3}
\Omega R^2\leq 2\alpha NR^4+\left(\beta-\frac{\epsilon}{C_*R^{2\sigma}}\right)\sum_{n\in\mathbb{Z}^N}|\phi_n|^{2\sigma+2}.
\end{eqnarray}
By using (\ref{com1h}), this time for $p=2\sigma+2$ and $q=2$, the term $\sum_{n\in\mathbb{Z}^N}|\phi_n|^{2\sigma+2}$ of (\ref{dd3}) can be estimated in terms of the power  $\sum_{n\in\mathbb{Z}^N}|\phi_n|^{2}=R^2$, as
\begin{eqnarray*}
\sum_{n\in\mathbb{Z}^N}|\phi_n|^{2\sigma+2}\leq \left(\sum_{n\in\mathbb{Z}^N}|\phi_n|^{2}\right)^{\frac{2\sigma+2}{2}}=R^{2\sigma+2}.
\end{eqnarray*}
Thus, from (\ref{dd3}) and the above estimate, we derive that
\begin{eqnarray*}
\Omega R^2\leq 2\alpha NR^4+\left(\beta-\frac{\epsilon}{C_*R^{2\sigma}}\right)R^{2\sigma+2},
\end{eqnarray*}
implying that the power satisfies the inequality
\begin{eqnarray}
 \label{dd6}
\left(\Omega+\frac{\epsilon}{C_*}\right)\leq 2\alpha NR^2+\beta R^{2\sigma}.
\end{eqnarray}
%%%%%%%%%%%%%%%%%%%%%%%%%%%%%%%%%%%%%%%%%%%%%%%%%%%%%%%%%%%%%%%%%%%%%%%%%%%%%%%%%%%%%%%%%%
%%%%%%%%%%%%%%%%%%%%%%%%%%%%%%%%%%%%%%%%%%%%%%%%%%%%%%%%%%%%%%%%%%%%%%%%%%%%%%%%%%%%%%%%%%
\begin{theorem}\label{notrih2}
Assume that $\sigma\geq 2/N$ and the parameters $\alpha, \beta,\Omega>0$
Let $\hat{R}_{\mathrm{crit}}>0$ denote the unique positive root of the polynomial equation
\begin{eqnarray*}
 2\alpha NR^2+\beta R^{2\sigma}-\left(\Omega+\frac{\epsilon}{C_*}\right)=0.
\end{eqnarray*}
Then a breather solution $\psi_n(t)=e^{\mathrm{i}\Omega t}\phi_n$, for any $\Omega>0$ of (\ref{hopDNLS}) must have power $\mathcal{P}>\hat{R}_{\mathrm{crit}}^2$.
\end{theorem}

For an even more explicit estimate, at least an estimation of the optimal constant $C_*$ is needed. This is provided by
\setcounter{proposition}{2}
\begin{proposition}
 \label{estimopt}
Let $\sigma\geq 2/N$. There exists $\nu_{\mathrm{crit}}>1/2$ such that the optimal constant of the inequality (\ref{WGN1}) satisfies
\begin{eqnarray}
\label{Cest}
 \frac{1}{4N}<C_*<\frac{\nu_{\mathrm{crit}}}{\sqrt{2\nu_{\mathrm{crit}}-1}}\dot\frac{2\sigma+1}{4N},\;\;N\geq 1.
\end{eqnarray}
 \end{proposition}
\textbf{Proof:} One of the fundamental results of \cite{Wein99} is the characterization of the optimal constant $C_*$
involving the \emph{excitation threshold} for breathers of the focusing DNLS equation with power nonlinearity. For instance it is known that
\begin{eqnarray*}
\mathcal{R}_{\mathrm{thresh}}=\left[\frac{(\sigma+1)\epsilon}{C_*}\right]^{\frac{1}{\sigma}},
\end{eqnarray*}
On the other hand, it was proved in \cite[Proposition II.1, pg. 6]{JFN2009}, that there exists $\nu_{\mathrm{crit}}>1/2$ such that
\begin{eqnarray}
\label{res1}
\left[\frac{\sqrt{2\nu_{\mathrm{crit}}-1}}{\nu_{\mathrm{crit}}}\cdot\frac{4N\epsilon(\sigma+1)}{2\sigma+1}
\right]^{\frac{1}{\sigma}}<R_{\mathrm{thresh}}<\left[4\epsilon N(\sigma+1)\right]^{\frac{1}{\sigma}}.
\end{eqnarray}
The estimate (\ref{Cest}) follows by inserting the characterization for $\mathcal{R}_{\mathrm{thresh}}$ into (\ref{res1}).  \ \ $\diamond$

Together with Proposition \ref{estimopt}, Theorem \ref{notrih2} can be restated and refined as follows.
\setcounter{theorem}{3}
\begin{theorem}
 \label{fintheo}
We assume that
\begin{eqnarray}
 \label{maxsigma}
\sigma\geq 2\;\;\mbox{when $N=1$}\;\;\mbox{and}\;\;\sigma> 1\;\;\mbox{when $N\geq 2$}.
\end{eqnarray}
Then a breather solution of (\ref{hopDNLS}) satisfies the lower bound
\begin{eqnarray}
\label{unspec}
 \left[\frac{1}{2\beta}\left(\Omega+\frac{4\epsilon N}{2\sigma+1}\dot \frac{\sqrt{2\nu_{\mathrm{crit}}-1}}{\nu_{\mathrm{crit}}}
-\frac{(2\alpha N)^{\frac{\sigma}{\sigma-1}}}{(\beta\sigma)^{\frac{1}{\sigma-1}}}\frac{\sigma-1}{\sigma}\right)\right]^{\frac{1}{\sigma}}<R^2,
\end{eqnarray}
in either the cases\newline
(i) (lattice spacing condition) For all $\Omega>0$ if
\begin{eqnarray}
 \label{mesh}
\epsilon>\frac{(2\alpha N)^{\frac{\sigma}{\sigma-1}}}{(\beta\sigma)^{\frac{1}{\sigma-1}}}\frac{(\sigma-1)(2\sigma+1)}{4N\sigma}
\frac{\nu_{\mathrm{crit}}}{\sqrt{2\nu_{\mathrm{crit}}-1}}.
\end{eqnarray}
(ii) (frequency condition) For all $\epsilon>0$ if
\begin{eqnarray}
 \label{freqmesh}
\Omega>\frac{(2\alpha N)^{\frac{\sigma}{\sigma-1}}}{(\beta\sigma)^{\frac{1}{\sigma-1}}}\frac{\sigma-1}{\sigma}.
\end{eqnarray}
\end{theorem}
\textbf{Proof:}. Inequality (\ref{dd6}) can be strengthened from below by replacing $1/C^*$ by its lower estimate as indicated from (\ref{Cest}). Then, (\ref{unspec}) comes out exactly as in Theorem \ref{posFreq} C.\ \ $\diamond$

We remark that in the case of the limit $a=0$, if we will repeat the calculations leading to the energy equation (\ref{dd1}) and inequalities (\ref{dd2})-(\ref{dd3}), we derive the inequality
\begin{eqnarray}
 \label{dd4}
0<\Omega R^2\leq\left(\beta-\frac{\epsilon}{C_*R^{2\sigma}}\right)\sum_{n\in\mathbb{Z}^N}|\phi_n|^{2\sigma+2}
\end{eqnarray}
Now, the positivity of the right-hand-side of (\ref{dd4}) implies that in the limit $\alpha=0$, the $\Omega$-independent lower bound
\begin{eqnarray*}
%\label{dd5}
\left[\frac{\epsilon}{C_*\beta}\right]^{\frac{1}{\sigma}}<R^2.
\end{eqnarray*}
is satisfied.

Let us also remark that the non-existence result of Theorem \ref{notrih} is valid in finite lattices, due to the validity of inequality (\ref{com1h}) in the subspace $\ell^2(\mathbb{Z}_N^K)$ of $\ell^2(\mathbb{Z}^N)$. Thus,  the result can be proved in the  case of finite lattices without any additional implications. Similarly, inequality (\ref{WGN1}) is also valid in
$\ell^2(\mathbb{Z}_N^K)$ and the estimates of Theorem \ref{fintheo} can be proved to be valid in finite lattices.  The estimates of Theorem \ref{fintheo} for the case $\sigma\geq 2/N$,  will be tested numerically in the next section.
%%%%%%e%%%%%%%%%%%%%%%5555
%%%%%%%%%%%%%%%%%%%%%%%%%%
%%%%%%%%%%%%%%%%%%%%%%%%%%
\section{Numerical study}
We present in this section, numerical results testing the behavior
and relevance of the theoretical estimates, in the case of the
$\mathrm{1D}$ lattice. The structure of this section has as follows.
In Sec. \ref{subs1} we analyze theoretically a refinement of the
original variational estimates on the example of the focusing case
$\alpha,\beta>0$, aiming to improve the capture of the contribution
of the linear part of the system to the power. This contribution is
manifested in the bounds, by the first eigenvalue of the discrete
Laplacian. The refinement takes into account the localization of
true breather solutions, by performing a ``cut-off'' procedure,
focusing on the most excited states. The improvement  is reflected
in the numerical simulations performed in Sec. \ref{subs2} for the
case of the cubic nonlinearity $\sigma=1$, showing in particular,
that in some cases of the weak coupling regime, the estimates
provide an accurate prediction of the numerical power. In Sec.
\ref{subs3} we present the numerical results for the case of
the quintic nonlinearity $\sigma=2$. The refined variational estimates
are valid, due to the translational invariance of the ``cut-off''
procedure, even in the case of the infinite lattices, and have been
tested against the interpolation estimates (e.g. those by the
interpolation inequality of Gagliardo-Nirenberg type). It was
interesting to observe that the refined variational bounds give a
better qualitative prediction when the nonlinearity parameter
$\beta$ is varied, while the interpolation estimates behave better
for large values of frequencies $\Omega$. Finally, in Sec.
\ref{subB}, we present an indicative numerical study of the
interpolation estimates in the defocusing case $\alpha<0$,
$\beta<0$. The main finding here is that the theoretical predictions
are improved for large values of the parameters $\beta$ and
$\sigma$.

We note that in all the numerical simulations,  the results have been obtained for a
$\mathrm{1D}$-lattice of $K=101$ particles.
%%%%%%%%%%%%%%%%%%%%%%%%%%%%%%%%%%%%%%%%%%%%%%%%%%%%%%%%%%%%%%%%%%%%%%%%%%%%%%%%%%%%%5555
%%%%%%%%%%%%%%%%%%%%%%%%%%%%%%%%%%%%%%%%%%%%%%%%%%%%%%%%%%%%%%%%%%%%%%%%%%%%%%%%%%%%%%%%
\setcounter{paragraph}{0}
\subsection{Focusing case
($\alpha>0,\beta>0$) with Dirichlet boundary conditions. Solutions
$\psi_n(t)=e^{\mathrm{i}\Omega t}\phi_n$, $\Omega>0$.}
\subsubsection{Theoretical analysis of the ``cut off'' procedure.}
\label{subs1}
According to
the results of Theorem \ref{posFreq} A., without any restrictions on
the exponent $\sigma>0$ of the nonlinearity, the first lower bound
comes from the positive root $R_{*,f}$ of the equation (\ref{eqR})
\begin{eqnarray}
\label{eqRf}
\beta\chi^{2\sigma}+2\alpha N\chi^2-(\mu_1+\Omega)=0,\;\;\sigma>0,\;\;N\geq 1
\end{eqnarray}
Then any breather solution has power $\mathcal{P}[\phi]$ satisfying the lower bound
\begin{eqnarray}
 \label{eqRfv1}
R_{*,f}^2<\mathcal{P}, \;\;\mbox{for all}\;\;\sigma>0,\;\;N\geq 1
\end{eqnarray}
In the particular case of the \textit{cubic nonlinearity} this lower bound reads as
\begin{eqnarray}
\label{cubicRegion1} \frac{\mu_1+\Omega}{2\alpha N
+\beta}<\mathcal{P},\;\;\sigma=1,\;\;N\geq 1.
\end{eqnarray}

Due to its relevance from a physical point of view, we have chosen
the cubic nonlinearity for a first numerical test.
The principal eigenvalue in (\ref{cubicRegion1}) manifests the contribution of the linear part of (\ref{hopDNLS1D}). The variational characterization of the principal eigenvalue (\ref{eigchar}), shows that the contribution of the linear part to the real breather is estimated from below by the eigenvector $\phi^1$ corresponding to the principal eigenvalue $\mu_1$, since the infimum in (\ref{eigchar}) is attained by $\phi^1$ as
\begin{eqnarray}
\label{eigatt}
\mu_1=\frac{(-\epsilon\Delta_d\phi^1,\phi^1)_{2}}{\sum_{|||n|||\leq K}|\phi^1_n|^2},
\end{eqnarray}
and (\ref{eigchar}) holds for all $\phi\in\ell^2(\mathbb{Z}^N_K)$. Qualitatively and geometrically, this approximation of the linear part seems reasonable, especially for breather solutions without sign changes (zero-crossings), since the eigenvector $\phi^1$ has no sign-changes. On the other hand, real simulations should consider a sufficiently large chain length $L$, especially when the infinite chain is modeled in order to avoid the influence of boundary conditions. In this case, $\mu_1\rightarrow 0$ (see (\ref{case1.1})-(\ref{case2.2})) and the contribution of this approximation becomes negligible. This can be explained physically, taking into account the fact that the real breather solution has a localization length $L_{\mathrm{loc}}< <L$ while the eigenvector is extended
through the entire chain length $L$. Proceeding further, since the contribution to the power outside the breather width $L_{\mathrm{loc}}$ is also negligible, we could ``cut-off'' the estimation procedure, estimating the power in $L_{\mathrm{loc}}$ and the contribution of the linear part by the principal eigenvalue $\mu_{1,L_{\mathrm{loc}}}$ of (\ref{DLap}) considered on $L_{\mathrm{loc}}$. Practically, since the breather width $L_{\mathrm{loc}}$ is unknown, we may perform this ``cut-off'' procedure in an interval close to the interval of unit length $L=1$, expecting that  the main contribution to the power comes from the excited sites included in the unit interval.  This is certainly true for breathers centered around the center of the interval $[-L,L]$ located at the site $n=\frac{(K+1)}{2}$. It should be remarked that a breather can be always centered around the 
principal site, especially in the infinite lattice due to the integer translation invariance therein.

For instance, we will  consider the interval $U=\left[-\frac{1}{2},\frac{1}{2}\right]$ together with the first neighbors adjacent to the points $-\frac{1}{2}$ and $\frac{1}{2}$. We assume that the breather configuration is described by the vector $\phi\in\mathbb{R}^{K+2}$
\begin{eqnarray}
\label{br1}
\phi=\left(\phi_0,\phi_1,\ldots\phi_{K+1}\right),\;\;\phi_0=\phi_{K+1}=0,
\end{eqnarray}
where $\phi_n:=\phi(x_n)$, $x_n=-\frac{L}{2}+nh$, $n=0,\ldots, K+1$. The number of oscillators located outside the piece of the chain of unit length
$U=\left[-\frac{1}{2},\frac{1}{2}\right]$ is
\begin{eqnarray}
\label{br2}
\theta=2\left\lceil\frac{\left(\frac{L}{2}-\frac{1}{2}\right)(K+1)}{L}\right\rceil,
\end{eqnarray}
where $\lceil x\rceil =\min\left\{n\in\mathbb{Z}\,|\,n\geq x\right\}$, $x\in\mathbb{R}$. Then the number of oscillators included in the unit interval $U$ is
\begin{eqnarray}
\label{br3}
m=K+2-\theta.
\end{eqnarray}
We also assume that the neighbors adjacent to the endpoints of $U$, are located at the sites $k$ and $k+m+1$. Note that these neighbors coincide with the end-points of $U$ only when $\frac{1}{h}\in\mathbb{N}$.  The distance $y\geq 0$  of these neighbors from the endpoints of $U$ is given by
\begin{eqnarray}
\label{br4a}
y&=&h-\frac{1-(m-1)h}{2},\;\;\mbox{if}\;\;\frac{1}{h}\notin\mathbb{N},\\
\label{br4b}
y&=&0,\;\;\mbox{if}\;\;\frac{1}{h}\in\mathbb{N}.
\end{eqnarray}
We denote by $U'$ the interval occupied by the $m$ oscillators in U and the two neighbors adjacent to the endpoints of $U$, i.e, containing $m+2$ oscillators. The length of $U'$ is
\begin{eqnarray}
\label{br5}
L'=1+2y.
\end{eqnarray}
We have the following
\begin{proposition}
\label{UL}
Let $\epsilon>0$, $N=\sigma=1$. Then the power of the $m$ oscillators included in the interval $U$
\begin{eqnarray*}
\mathcal{P}_U=\sum_{n=k+1}^{k+m}|\phi_n|^2,
\end{eqnarray*}
satisfies the estimate
\begin{eqnarray}
\label{cRU} \frac{4\epsilon\sin^2\left(\frac{\pi}{2(m+1)}\right)+\Omega}{2\alpha
+\beta}<\mathcal{P}_U<\mathcal{P},
\end{eqnarray}
where the number of points $m$ in $U$ is given by (\ref{br3}).
\end{proposition}
{\textbf{Proof:} The breather configuration vector $\phi$ in (\ref{br1}) can be decomposed as $\phi=\phi^{L\setminus U}+\phi^{U}$
\begin{eqnarray}
\label{br7}
\phi^{L\setminus U}&=&\left(\phi_0,\phi_1,\ldots\phi_{k-1},\phi_{k},0,\ldots,0,\phi_{k+m+1},\ldots,\phi_{K+2}\right),\\
\label{br8}
\phi^{U}&=&\left(0,\ldots,0,\phi_{k+1},\phi_{k+2},\ldots,\phi_{k+m},0\ldots,0\right).
\end{eqnarray}
Since the decomposition is linear, at first glance the elements $\phi^{L\setminus U}$ and $\phi^U$ satisfy the equations
\begin{eqnarray*}
-\epsilon\Delta_d\phi^U_n+\Omega\phi^U_n
-\alpha\phi^U_n\left(|\phi_{n+1}|^2
+|\phi_{n-1}|^2\right)+\beta|\phi_n|^{2\sigma}\phi^U_n=0,\;\;n=0,\ldots,K+2,\\
-\epsilon\Delta\phi^{L\setminus U}_n+\Omega\phi^{L\setminus U}_n
-\alpha\phi^{L\setminus U}_n\left(|\phi_{n+1}|^2
+|\phi_{n-1}|^2\right)+\beta|\phi_n|^{2\sigma}\phi^{L\setminus U}_n=0,\;\;n=0,\ldots,K+2.
\end{eqnarray*}
However, on the account of (\ref{br8}), the equation for $\phi_U$ can be written as
\begin{eqnarray*}
-\epsilon\Delta_d\phi^U_n&+&\Omega\phi^U_n
-\alpha\phi^U_n\left(|\phi^U_{n+1}|^2
+|\phi^U_{n-1}|^2\right)+\beta|\phi^U_n|^{2\sigma}\phi^U_n=0,\;\;n=k+1,\ldots,k+m+1,\\
\phi^U_{k}&=&\phi^U_{k+m+1}=0.
\end{eqnarray*}
Relabeling for convenience, the system for $\phi^U$ can be considered on the interval $U'$ of the $m+2$ oscillators $j=0,\ldots,m+2$ as
\begin{eqnarray}
\label{br9A}
-\epsilon\Delta_d\phi^U_j&+&\Omega\phi^U_j
-\alpha\phi^U_j\left(|\phi^U_{j+1}|^2
+|\phi^U_{j-1}|^2\right)+\beta|\phi^U_j|^{2\sigma}\phi^U_j=0,\;\;j=1,\ldots,m,\\
\label{br9B}
\phi^U_{0}&=&\phi^U_{m+1}=0.
\end{eqnarray}
We also consider the linear eigenvalue problem on $U'$
\begin{eqnarray}
\label{br10A}
-\epsilon\Delta_d\phi_j&=&\mu\phi_j,\;\;j=1,\ldots,m,\\
\label{br10B}
\phi_{0}&=&\phi_{m+1}=0.
\end{eqnarray}
The principal eigenvalue $\mu_{1,U'}$ of (\ref{br10A})-(\ref{br10B}) is given by
\begin{eqnarray}
\label{br11}
\mu_{1,U'}=4\epsilon\sin^2\left(\frac{\pi L'}{2L'(m+1)}\right)=4\epsilon\sin^2\left(\frac{\pi}{2(m+1)}\right).
\end{eqnarray}
Repeating the calculations of the proof of Theorem \ref{posFreq} on the system (\ref{br9A})-(\ref{br9B}), we derive that
\begin{eqnarray*}
 \frac{\mu_{1,U'}+\Omega}{2\alpha
+\beta}<\mathcal{P}_U<\mathcal{P},\;\;\sigma=1,\;\;N\geq 1,
\end{eqnarray*}
i.e, the left-hand side of (\ref{cRU}). The left hand side follows from the fact that $\mathcal{P}=\sum_{j=1}^{K+2}|\phi_j|^2>P_U$.\ \ $\diamond$
\setcounter{remark}{1}
\begin{remark}
\label{remU}
{\em The estimate (\ref{cRU}) will be useful for the numerical simulations since it is valid for any $\epsilon$ and can be used for the fully discrete case, even in the case of an infinite lattice, since the interval $U'$ where the procedure takes place, is the same independently of the length of the chain. Thus even in the case  $h>0.5$, where the unit interval $U$ contains only the centered site, we may perform the ``cut-off'' procedure for the centered site and the two adjacent neighbors occupying the interval $U'$ of length $L'=1+2y=1+2(h-\frac{1}{2})$.  The estimate (\ref{cRU}) reads as
\begin{eqnarray}
\label{cRUA} \frac{4\epsilon\sin^2\left(\frac{\pi}{4}\right)+\Omega}{2\alpha
+\beta}<\mathcal{P}_U<\mathcal{P},
\end{eqnarray}
estimating the power of the breather in terms of the ``most excited site''. }
\end{remark}

In the case we approximate the continuous limit by considering $\epsilon>0$ sufficiently large, we have
\setcounter{proposition}{2}
\begin{proposition}
\label{ULC}
Let $\epsilon\cong\frac{1}{h^2}$, $N=\sigma=1$. Assume that $\epsilon$ is sufficiently large, or $\frac{1}{h}\in\mathbb{N}$. The power of the $m$ oscillators included in the interval $U$
\begin{eqnarray*}
\mathcal{P}_U=\sum_{n=k+1}^{k+m}|\phi_n|^2,
\end{eqnarray*}
satisfies
\begin{eqnarray}
\label{cRUhs} \frac{4{(m+1)^2}\sin^2\left(\frac{\pi}{2(m+1)}\right)+\Omega}{2\alpha
+\beta}<\mathcal{P}_U<\mathcal{P},
\end{eqnarray}
where the number of points $m$ in $U$ is given by (\ref{br3}).
\end{proposition}
\textbf{Proof:} Working as in the proof of Proposition \ref{UL}, we estimate the linear part of (\ref{br9A})-(\ref{br9B}),  by using the principal eigenvalue of the linear problem (\ref{br10A})-(\ref{br10B}), where in the case $\epsilon\cong\frac{1}{h^2}$, is
\begin{eqnarray}
\label{br12}
\mu_{1,U'}\cong\frac{4}{h'^2}\sin^2\left(\frac{\pi h'}{2L'}\right)=4\frac{(m+1)^2}{(1+2y)^2}\sin^2\left(\frac{\pi}{2(m+1)}\right),
\end{eqnarray}
since the spacing of $U'$ is
\begin{eqnarray*}
h'=\frac{L'}{m+1}=\frac{1+2y}{m+1}.
\end{eqnarray*}
The distance $y$ is defined in (\ref{br4a})-(\ref{br4b}). Letting $h\rightarrow 0$ we have $y\rightarrow 0$ (not monotonically), and (\ref{br12}) implies that
\begin{eqnarray}
\label{br13}
\mu_{1,U'}\cong 4(m+1)^2\sin^2\left(\frac{\pi}{2(m+1)}\right).
\end{eqnarray}
When $\frac{1}{h}\in\mathbb{N}$, the end-points of $U'$ are $x_k=-1/2$ and $x_{k+m+1}=1/2$, and $y=0$.\ \ $\diamond$

\setcounter{remark}{3}
\begin{remark}
\label{remUh}
{\em  When we approximate the continuum by considering $\epsilon>0$ sufficiently large, we observe that the principal eigenvalue $\mu_{1,U'}$ has the expression (\ref{case1.1}) for $L=1$, in terms of the number $m+2$ of oscillators occupying the interval $U'$. Clearly, since $\frac{1}{m+1}<1$ for $m\geq 1$
\begin{eqnarray}
\label{br14}
4<\mu_{1,U'}=4(m+1)^2\sin^2\left(\frac{\pi}{2(m+1)}\right)<\pi^2.
\end{eqnarray}
and we have the bounds
\begin{eqnarray}
\label{cRUhA}
\frac{4+\Omega}{2\alpha
+\beta} <\frac{\mu_{1,U'}+\Omega}{2\alpha
+\beta}<\mathcal{P}_U<\mathcal{P}.
\end{eqnarray}
Besides, for large $\epsilon>0$, $m>1$ is large enough and  (\ref{br14}) and (\ref{cRUhA}) justify the approximation
\begin{eqnarray*}
\mu_{1,U'}\sim 4\epsilon\sin^2\left(\frac{\pi}{2\sqrt{\epsilon}}\right),
\end{eqnarray*}
and the estimation of the power as
\begin{eqnarray}
\label{br15}
\frac{4+\Omega}{2\alpha
+\beta} <\frac{\mu_{1,U'}+\Omega}{2\alpha
+\beta}<\mathcal{P}_U<\mathcal{P},\;\;\mu_{1,U'}\sim 4\epsilon\sin^2\left(\frac{\pi}{2\sqrt{\epsilon}}\right).
\end{eqnarray}
}
\end{remark}

\subsubsection{Numerical results: cubic nonlinearity $\sigma=1$}
\label{subs2} We now turn to the presentation of the numerical
results which starts with the case $\epsilon=1$. The ``cut-off''
approximation of Proposition \ref{UL} takes place on the interval
$U'$ of length $L'=2$ ($y=0.5$) and the unit interval $U$ contains
only one site ($m=1$). In Fig. \ref{figeps1}(a), the real power of a
breather family is plotted using dots against the nonlinear
parameter $\alpha$. The lower bound obtained with the "cut-off"
procedure (\ref{cRUA}) is shown with  a triangle (grey) line.
Notice that it is always below the real power. The qualitative
prediction of the pattern of the numerical power as given by the
theoretical estimate should be remarked, due to the effective
approximation of the contribution of the linear and the nonlinear
part to the power.

The continuous approximation (\ref{br15}) in the unit length,
plotted with a continuous blue curve, is not satisfied as a lower
bound for all the values of the parameter $\alpha$ as expected,
since we are fairly far from the continuum limit. Remarkably,
however, we observe that for a quite large regime of the parameter
$\alpha$, the corresponding prediction is below the numerical power.
This is due to the fact that $\epsilon=1$ is a critical value for
our approximation in the sense that for $\epsilon=1$ the eigenvalue
$\mu_{1,U'}$ in (\ref{br15}) attains its minimum $\mu_{1,U'}=4$.

In Fig. \ref{figeps1}(b) the breather profile (continuous (red)
curve) is plotted against the eigenvector on $U'$ and the eigenvector on the
length of the chain $L$ for $\epsilon=\Omega=1$ and
$\beta=2$. Notice that the eigenvector in the length of the system
$L$ is spread out along the chain (on the scale of the figure it is
almost a horizontal  line) and its contribution to the estimates
would be negligible.
%%%%%%%%%%%%%%%%%%%%%%%%%%%%%%%%%%%%%%%%%%%%%%%%%%%%%%%%%%%%%%%%%%%%%%%%%%%%%%%%%%
\begin{figure}
\begin{center}
    \begin{tabular}{cc}
    \includegraphics[scale=0.69]{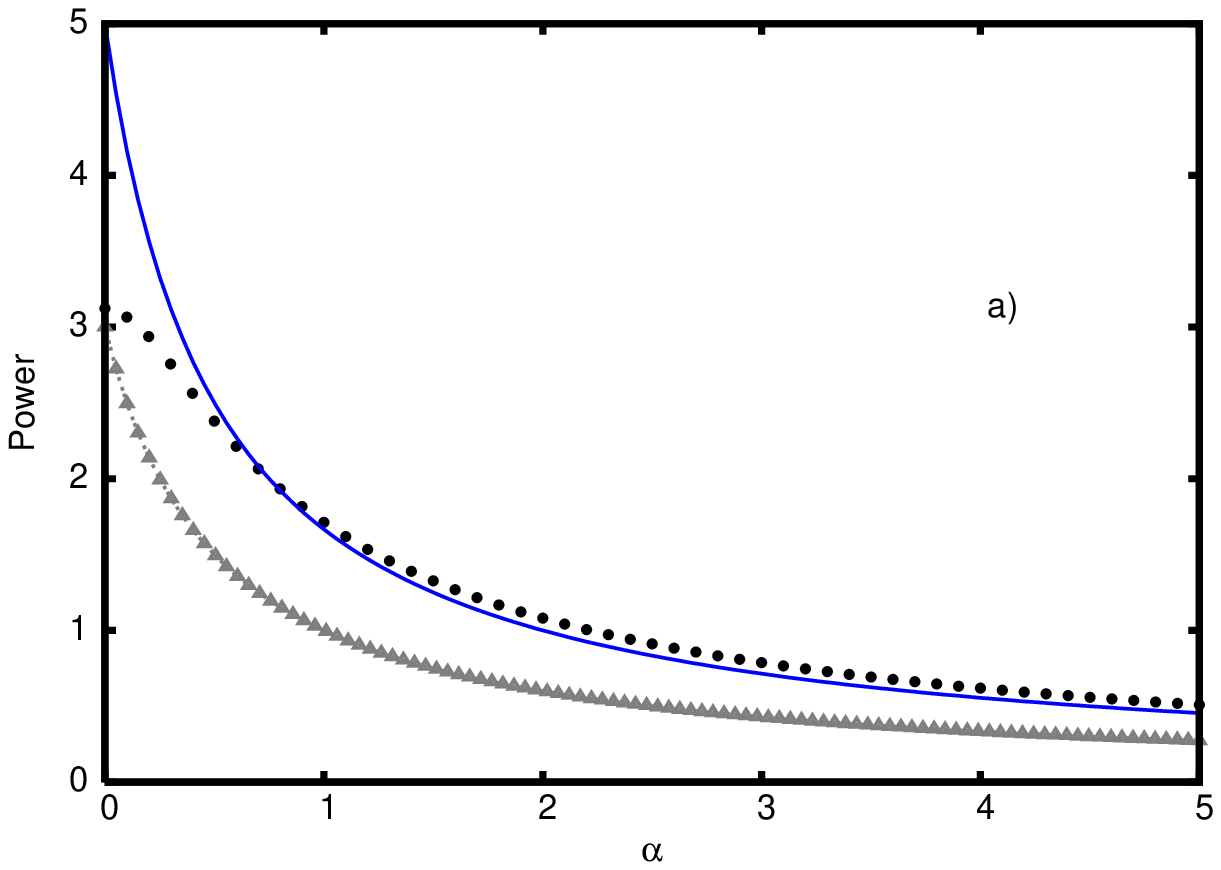} &
    \includegraphics[scale=0.69]{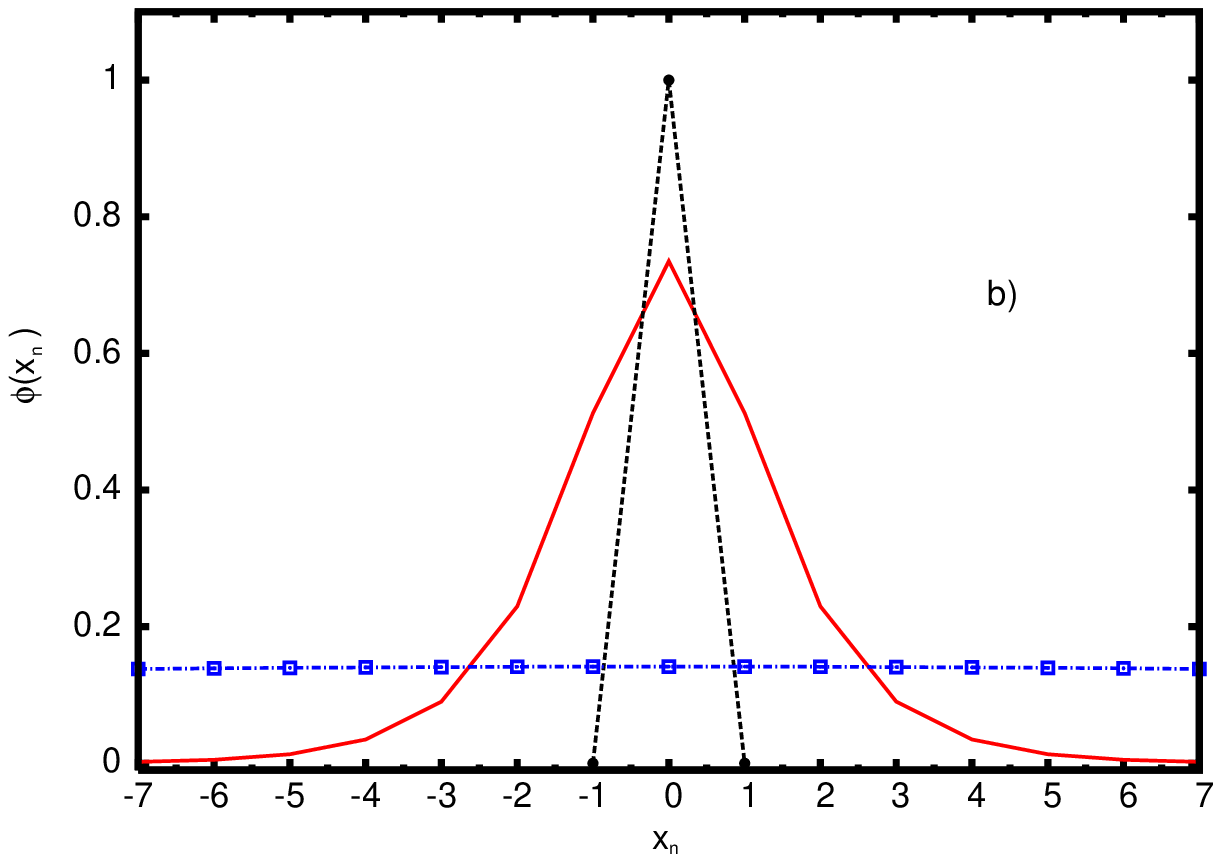}
    \end{tabular}
\caption{(a) Power of breathers versus nonlinear parameter $\alpha$
in the HDNLS system with cubic nonlinearity ($\sigma=1$) and $\epsilon=1$. Symbols (dots)
correspond to numerical calculations while the triangles (grey) line
represents estimation (\ref{cRUA}). The continuous (blue) curve corresponds to the estimate (\ref{br15}). Other parameters:
$\beta=\Omega=1$. (b) Breather profile (continuous (red) curve) against the eigenvector (dashed (black) curve) of (\ref{br10A})-(\ref{br10B}) on the  interval $U'$ of length $L'=2$. The eigenvector of (\ref{DLap}) in the length $L$ of the chain is represented by the dashed-boxes (blue) curve.}
\label{figeps1}
\end{center}
\end{figure}
%%%%%%%%%%%%%%%%%%%%%%%%%%%%%%%%%%%%%%%%%%%%%%%%%%%%%%%%%%%%%%%%%%%%%%%%%%%%%%%5555
%\newline

In Fig. \ref{figeps2} we present the results of the study for
$\epsilon=2$.  Triangles (grey curve) correspond again to the
estimate (\ref{cRUA}), still valid in the interval $U'$ having now
length $L'\sim 1.414$ and the unit interval $U$ contains one site
($m=1$). We observe the increased quantitative accuracy of the
prediction of the actual power (symbols (dots)). The continuous
approximation (\ref{br15}) in the unit length  represented by the
dash-dotted (blue) curve is not satisfied as a lower bound as
predicted by Propositions \ref{UL} and \ref{ULC}. Nevertheless, it
is worth observing that the continuous approximation is only
slightly above the actual value. This is connected to the fact that
increasing values of $\epsilon$ correspond to a closer approximation
of the continuous limit. The dotted (green) curve below the
triangles represents the initial estimate (\ref{cubicRegion1}) with
the eigenvalue $\mu_1$ corresponding to the eigenvector of
(\ref{DLap}) over the original length $L$ of the system. In this
case, the estimation of the contribution of the linear part to the
power is negligible as (\ref{case2.2}) shows, thus
(\ref{cubicRegion1}) is well below the actual power.
%%%%%%%%%%%%%%%%%%%%%%%%%%%%%%%%%%%%%%%%%%%%%%%%%%%%%%%%%%%%%%%%%%%%%%%%%%%%%%%%%%
\begin{figure}
\begin{center}
    \begin{tabular}{cc}
    \includegraphics[scale=0.69]{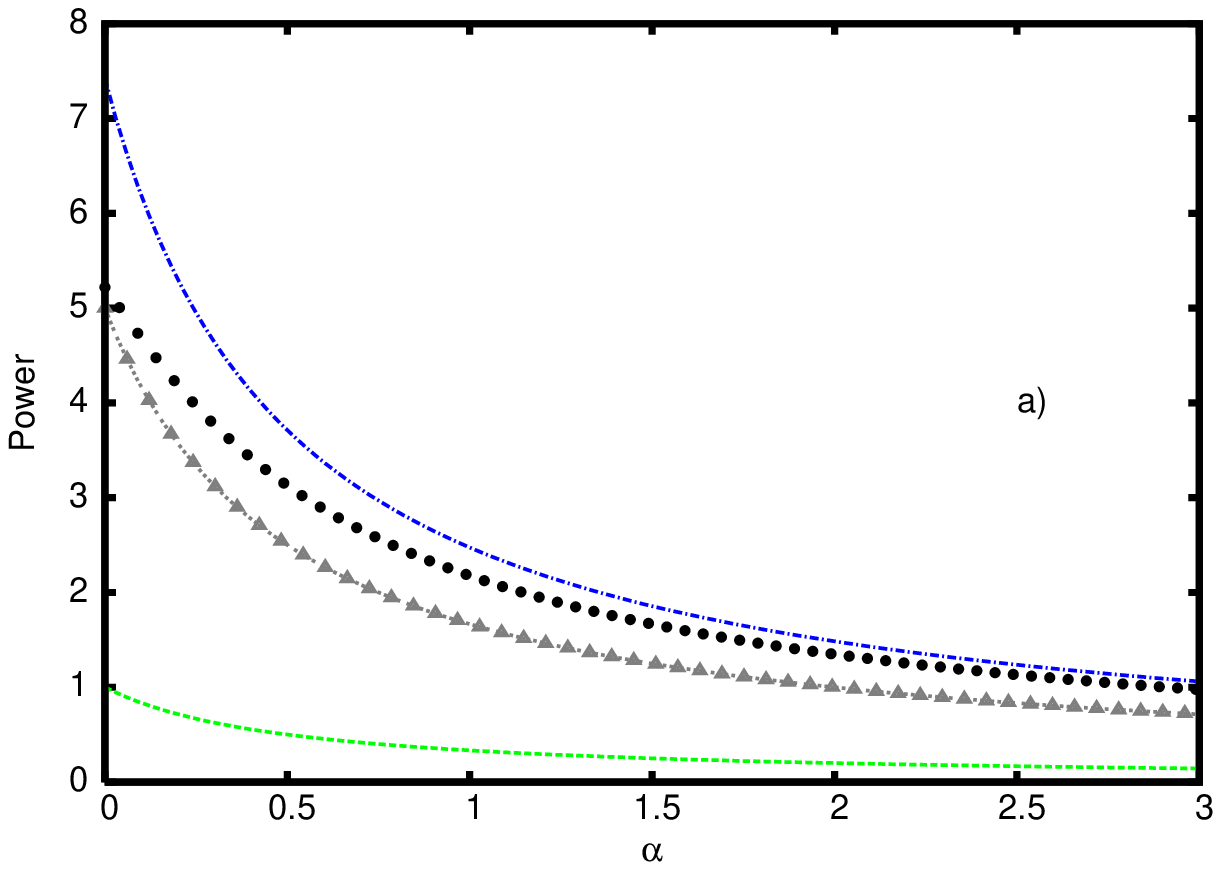} &
    \includegraphics[scale=0.69]{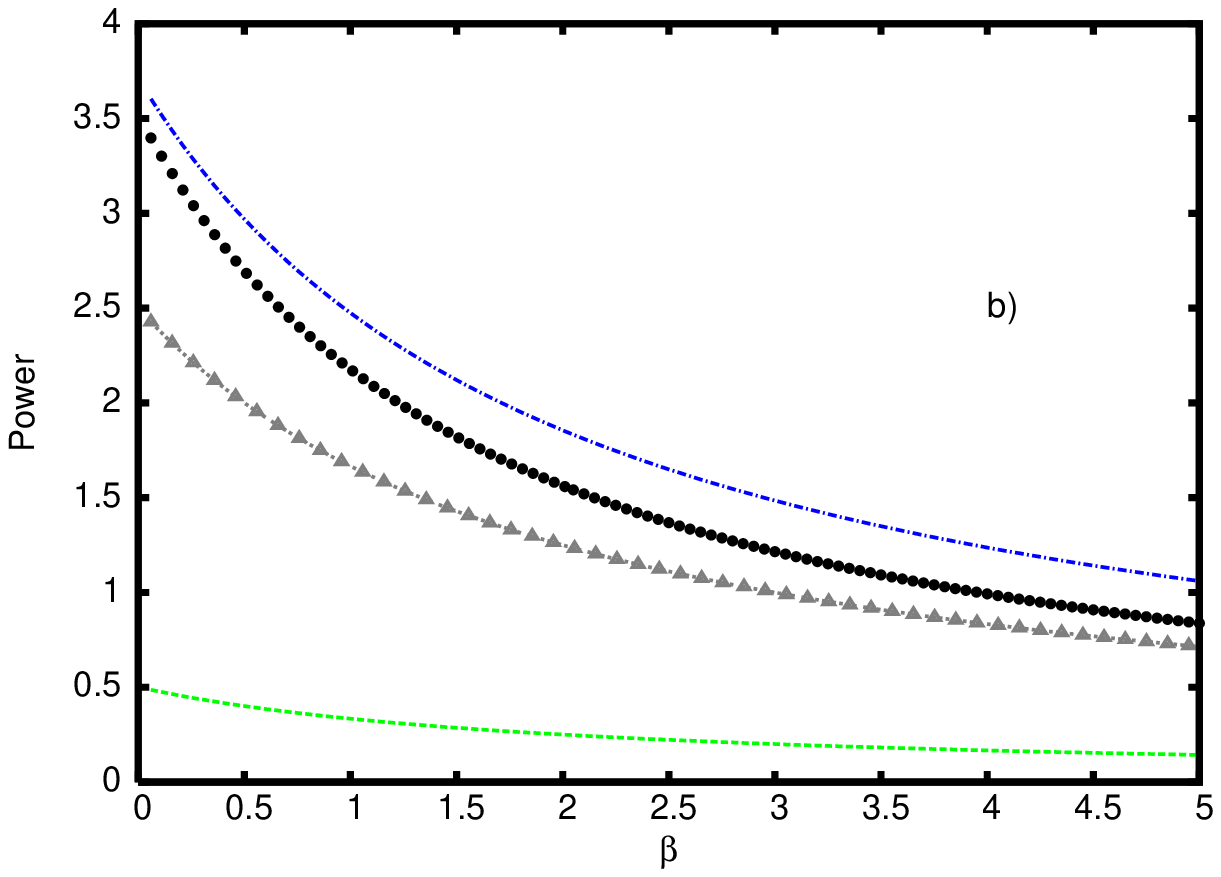}
    \end{tabular}
\caption{(a) Power of breathers versus nonlinear parameter $\alpha$
in the HDNLS system with cubic nonlinearity ($\sigma=1$) and
$\epsilon=2$. Symbols (dots) correspond to numerical calculations
while the triangles (grey) line represents estimation (\ref{cRUA})
obtained with the the ``cut-off'' approximation of Proposition
\ref{UL}. The dash-dotted (blue) curve corresponds to the
continuous approximation (\ref{br15}). The first dotted (green)
curve from below represents the initial estimate
(\ref{cubicRegion1}) with the eigenvalue $\mu_1$ calculated over the
length $L$ of the system. Other parameters: $\beta=\Omega=1$. (b)
The power and its estimates versus the nonlinear parameter $\beta$.
Other parameters are chosen as $\alpha=\Omega=1$.} \label{figeps2}
\end{center}
\end{figure}
%%%%%%%%%%%%%%%%%%%%%%%%%%%%%%%%%%%%%%%%%%%%%%%%%%%%%%%%%%%%%%%%%%%%%%%%%%%%%%%5555
%\newline

The effectiveness of the ``cut-off'' approximation of Proposition
\ref{UL} and Remark \ref{remU} on length $L'$, if compared with the
initial estimate (\ref{cubicRegion1}) on the length of the system
$L$ is even more transparent in the study for $\epsilon=3$, where
the results are presented in Fig. \ref{figeps3}. In this case
$L'\sim 1.154$ and still $m=1$.  The curves are traced as in Fig.
\ref{figeps2}, except the new continuous (red) curve which is above
the theoretical estimate (\ref{cubicRegion1}). This curve
corresponds to the lower bound in the left-hand side of
(\ref{br15}). We observe that the prediction of (\ref{cRUA}) is of
excellent accuracy throughout the continuation over the nonlinear
parameter $\alpha$ and of very good accuracy even versus the
nonlinear parameter $\beta$, being saturated for large values of
$\beta$. It seems that the theoretical estimates capture better the
variation over the nonlinear coupling coefficient $\alpha$ rather
than the onsite nonlinearity coefficient $\beta$. This is due to the
fact that through the estimation process of Theorem \ref{posFreq}
the contribution of the hopping nonlinearity is ``doubled'' by the
nonlinear coupling with the adjacent sites (see the inequality
(\ref{LB})), although both nonlinearities are of cubic order in the
case $\sigma=1$. For large values of $\beta$ the manifestation of
the power nonlinearity is stronger. More precisely, observe in
Fig.\ref{figeps3}(b) that the convergence of (\ref{cRUA}) to the
real power starts after $\beta\geq 2$, i.e. after ``doubling'' the
strength of the onsite nonlinearity.
%Since $L'\sim 1.154$ and $\epsilon$ has been increased further,
In this case, the continuous
approximation over the unit length approaches further the actual power
(still, however, from above).
%especially versus the nonlinear hopping parameter $\alpha$.

The approximation procedure considers the cases  $\epsilon=1,2,3,4$,
as  weak coupling cases, in the sense that the unit length $U$
contains only one point and the spacing is  $h>0.5$. Note that $U'$
has different length ($L'=2$ for $\epsilon=1$, $L'\sim 1.414$ for
$\epsilon=2$, $L'\sim 1.154$ for $\epsilon=3$ and $L'=1$ for
$\epsilon=4$.) Since $m=1$, the eigenvalue in $U'$  given in
(\ref{br11}) is always $\mu_{1,U'}=2\epsilon$. Thus, in the weak
coupling case, the continuous approximation in the unit length
(\ref{br15}) is not valid and (\ref{br15}) is not satisfied as a
lower bound for the power. On the other hand, the discrete
approximation with the cutoff procedure within (\ref{cRUA}) becomes
progressively better as $\epsilon$ is increased.
%%%%%%%%%%%%%%%%%%%%%%%%%%%%%%%%%%%%%%%%%%%%%%%%%%%%%%%%%%%%%%%%%%%%%%%%%%%%%%%%%%
\begin{figure}
\begin{center}
    \begin{tabular}{cc}
    \includegraphics[scale=0.69]{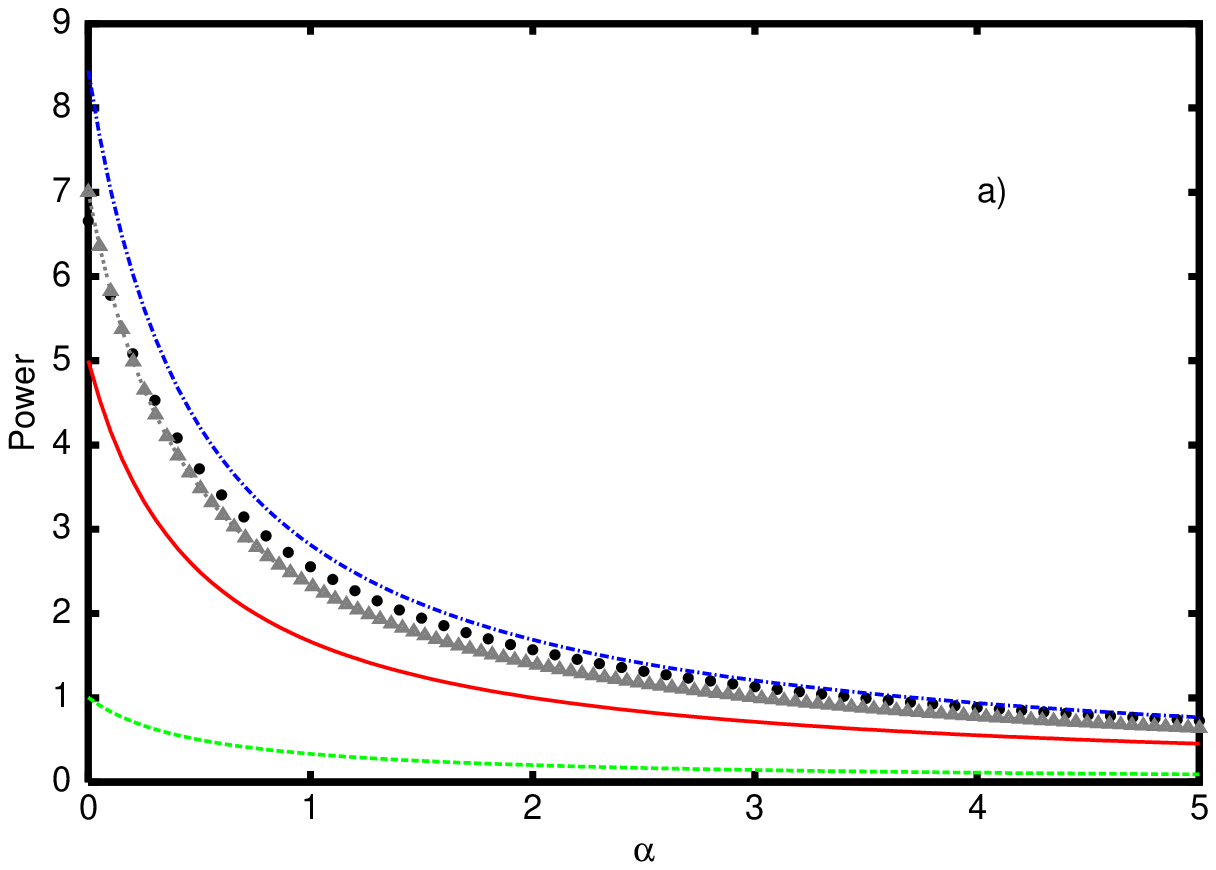} &
    \includegraphics[scale=0.69]{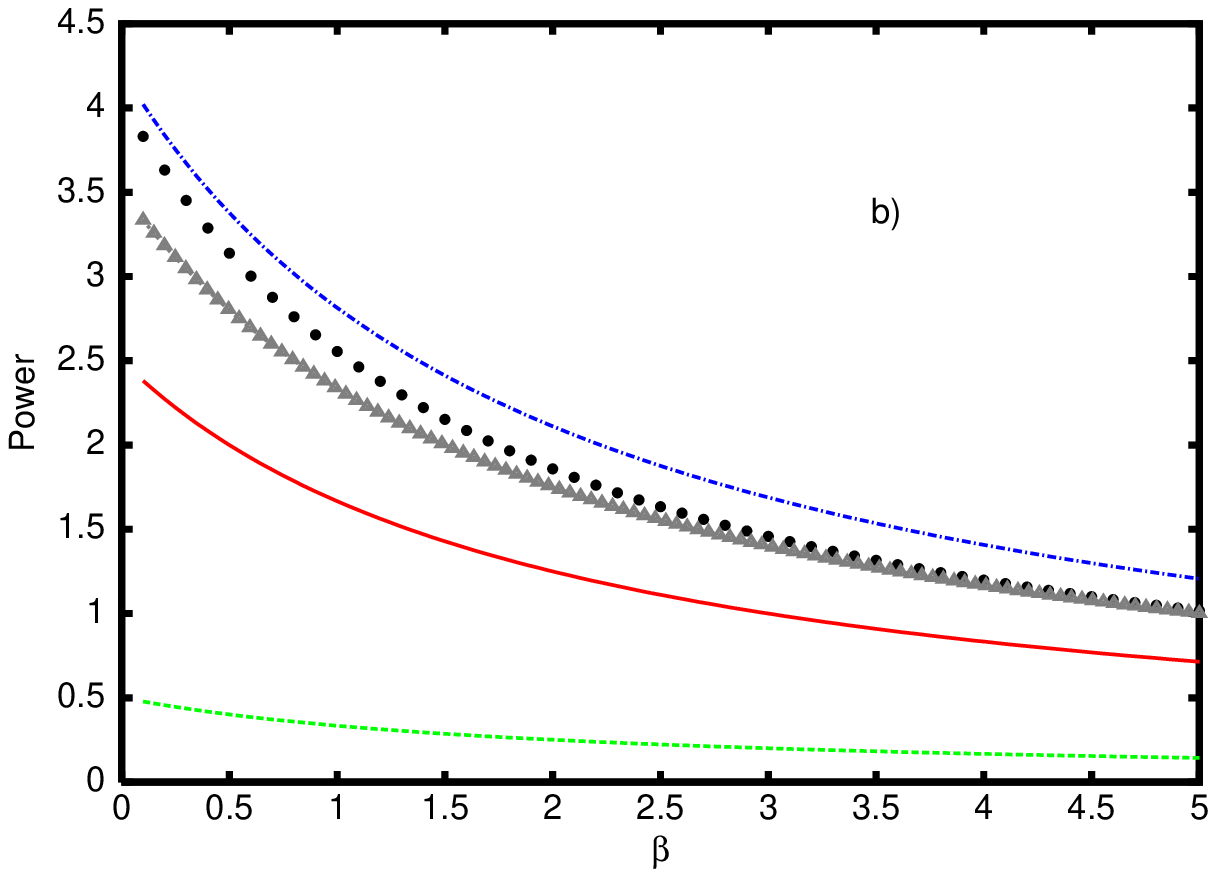}
    \end{tabular}
\caption{(a) Power of breathers versus the nonlinear parameter
$\alpha$ in the DNLS system with cubic nonlinearity ($\sigma=1$) and
$\epsilon=3$. Symbols (dots) correspond to numerical calculations
while the triangles (grey) line represents estimation (\ref{cRUA})
obtained with the  ``cut-off'' approximation of Proposition
\ref{UL}. The dash-dotted (blue) curve corresponds to the continuous
approximation (\ref{br15}). The first dotted (green) curve from
below represents the initial estimate (\ref{cubicRegion1}) with the
eigenvalue $\mu_1$ calculated over the length $L$ of the system.
Other parameters are chosen as $\beta=\Omega=1$. (b) The power and
its estimates versus the nonlinear parameter $\beta$. Other
parameters are $\alpha=\Omega=1$.} \label{figeps3}
\end{center}
\end{figure}
%%%%%%%%%%%%%%%%%%%%%%%%%%%%%%%%%%%%%%%%%%%%%%%%%%%%%%%%%%%%%%%%%%%%%%%%%%%%%%%%%%%%%%%%%%%%%%%%%%%%%%%%%%%%%%%%%%%%%%%%%%%%%%%%%%%%%%%%%

Propositions \ref{UL} and \ref{ULC} predict that the position of the
curves (\ref{cRUA}) and (\ref{br15}) should be interchanged when
$\epsilon>4$ ($h<0.5$). In this case, the unit interval $U$ contains
more than one site ($m>1$) and (\ref{cRUA}) is not valid. In Fig.
\ref{figeps10} we present the numerical study for $\epsilon=10$.
Here $U'$ has length $L'\sim 1.264$, the unit interval $U$ contains
three sites ($m=3$) and $h\sim 0.316$, which can be considered as approaching
the continuous limit. Note that for $\epsilon>4$ we have $L'\ge
1$, however $y\rightarrow 0$ as  $\epsilon$ is increased. We observe that
(\ref{cRUA}) is well above the actual breather power in this case,
while  now (\ref{br15}) provides an adequate approximation especially
versus the hopping parameter
$\alpha$.

Figure \ref{figeps1}(b) and Figures~\ref{figprofs3_10}(a)-(b) are
showing the breather profiles versus the eigenvectors on $U'$ and
the length $L$ of the system, and demonstrate the main features of
the approximation procedure. A first important feature is that both
the real breather and the approximating eigenvector for the linear
part contribution on $U'$ are localized. This is in contrast to
the eigenvector associated with $\mu_1$ (of the original problem)
which is extended over the entire length $L$ of the system. This
approximation of the linear part is effective for values of the weak
coupling, where the eigenvector on $U'$ has width comparable with
the localization length of the breather. In the anticontinuous
limit, we expect strong localization effects while the eigenvalue
$\mu_{1,U'}=2\epsilon$ becomes negligible again, and the estimates
are less effective.

The second feature is that although we are calculating only the
contribution to the energy of the sites included in $U'$, the
approximation is focusing on these sites being the principal excited
ones. Furthermore, Figures \ref{figeps1}(b) and Figures
\ref{figprofs3_10} (a),(b) demonstrate a concentration of the
``missing'' power of the sites outside $U'$ to the most excited
sites within the eigenvector on $U'$. This is observable by a
comparison of the profiles for $\epsilon=2,3,10$.
%, showing that when the
%peak is most excited, the better the approximation works ($\epsilon=3$).
From the strong coupling to the anticontinuous limit, the breather profile
approaches the continuous one, while the eigenvector in $U'$ converges to the
continuous eigenfunction. Then, both the difference between the breather and
the eigenvector width as well the difference of their ``peaks'' becomes
constant, and again, the estimates are becoming less effective. Besides, the
methods of this paper are making use of the properties of the discrete phase
space and should be extended appropriately in function spaces in order to
capture effectively the behavior of the continuous counterpart.
Nevertheless, in this setting of larger $\epsilon$, the continuum variant of
the approximation over
the interval $U'$ yields a suitable lower threshold for the breather
power.

%%%%%%%%%%%%%%%%%%%%%%%%%%%%%%%%%%%%%%%%%%%%%%%%%%%%%%%%%%%%%%%%%%%%%%%%%%%%%%%%%%%%%%%%%%%%%%%%%%%%%%%%%5555
\begin{figure}
\begin{center}
    \begin{tabular}{cc}
    \includegraphics[scale=0.69]{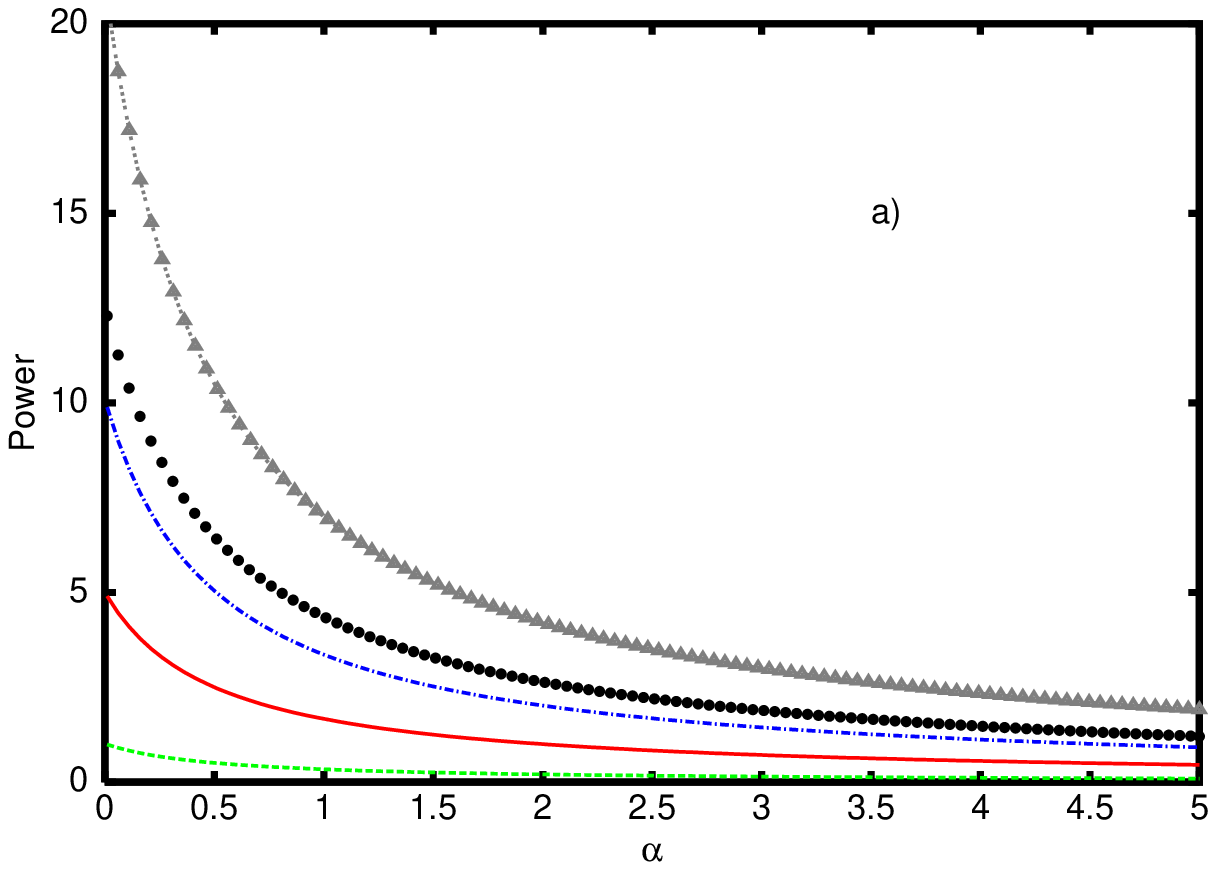} &
    \includegraphics[scale=0.69]{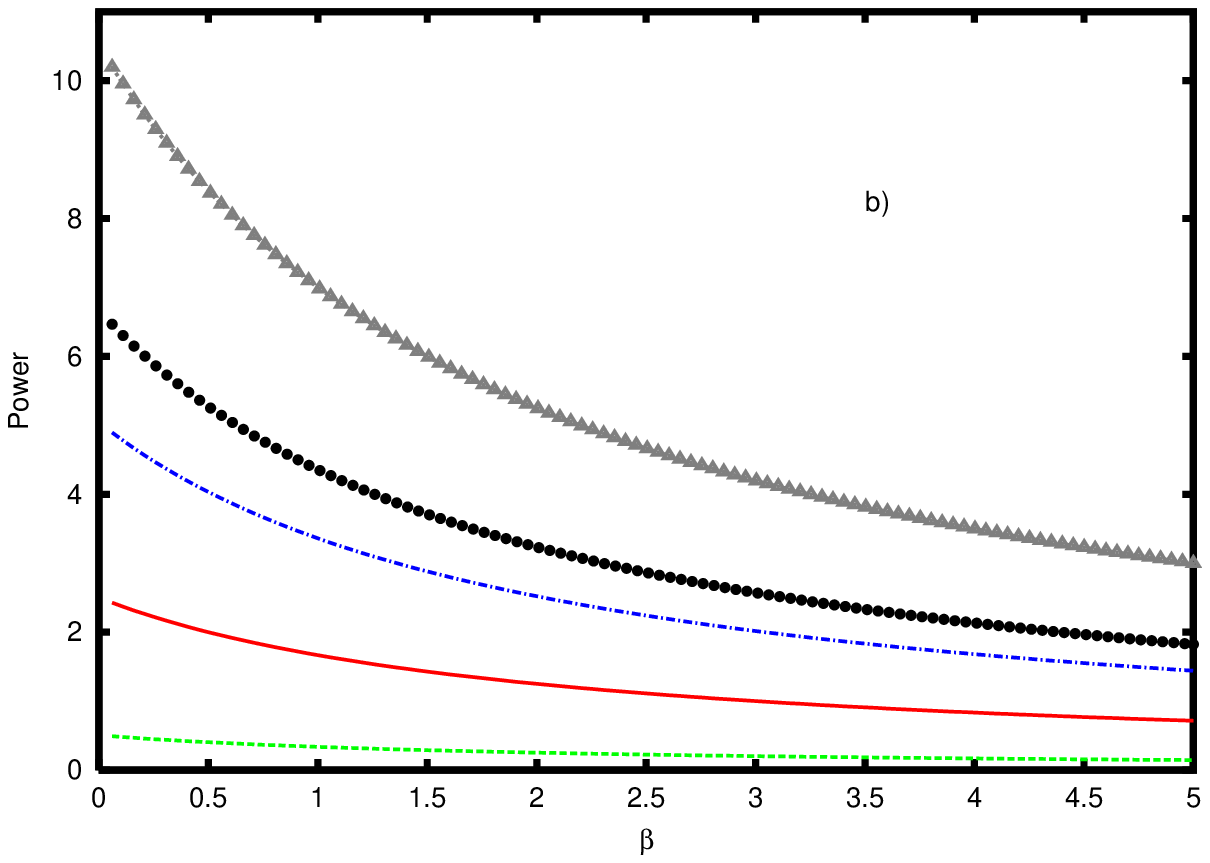}
    \end{tabular}
\caption{(a) Power of breathers versus nonlinear parameter $\alpha$
in the HDNLS system with cubic nonlinearity ($\sigma=1$) and
$\epsilon=10$. Symbols (dots) correspond to numerical calculations
while the triangles (grey) line represents estimation (\ref{cRUA})
obtained with the the ``cut-off'' approximation of Proposition
\ref{UL}. The dotted-dashed (blue) curve corresponds to the estimate
(\ref{br15}).  The first dotted (green) curve from below represents
the initial estimate (\ref{cubicRegion1}) with the eigenvalue
$\mu_1$ calculated in the length $L$ of the system and the
continuous (red) curve above stands for the estimate (\ref{br15})
with the lower bound $4\epsilon\leq\mu_1(\epsilon)$.  Other
parameters are chosen as $\beta=\Omega=1$. (b) The power and its
estimates are shown versus the nonlinear parameter $\beta$. Other parameters
$\alpha=\Omega=1$.} \label{figeps10}
\end{center}
\end{figure}
%%%%%%%%%%%%%%%%%%%%%%%%%%%%%%%%%%%%%%%%%%%%%%%%%%%%%%%%%%%%%%%%%%%%%%%%%%%%%%%%%%%%%%%%%%%%%%%%%%%%%%%%%%%%%%%%%
%%%%%%%%%%%%%%%%%%%%%%%%%%%%%%%%%%%%%%%%%%%%%%%%%%%%%%%%%%%%%%%%%%%%%%%%%%%%%%%%%%
\begin{figure}
\begin{center}
    \begin{tabular}{cc}
    \includegraphics[scale=0.69]{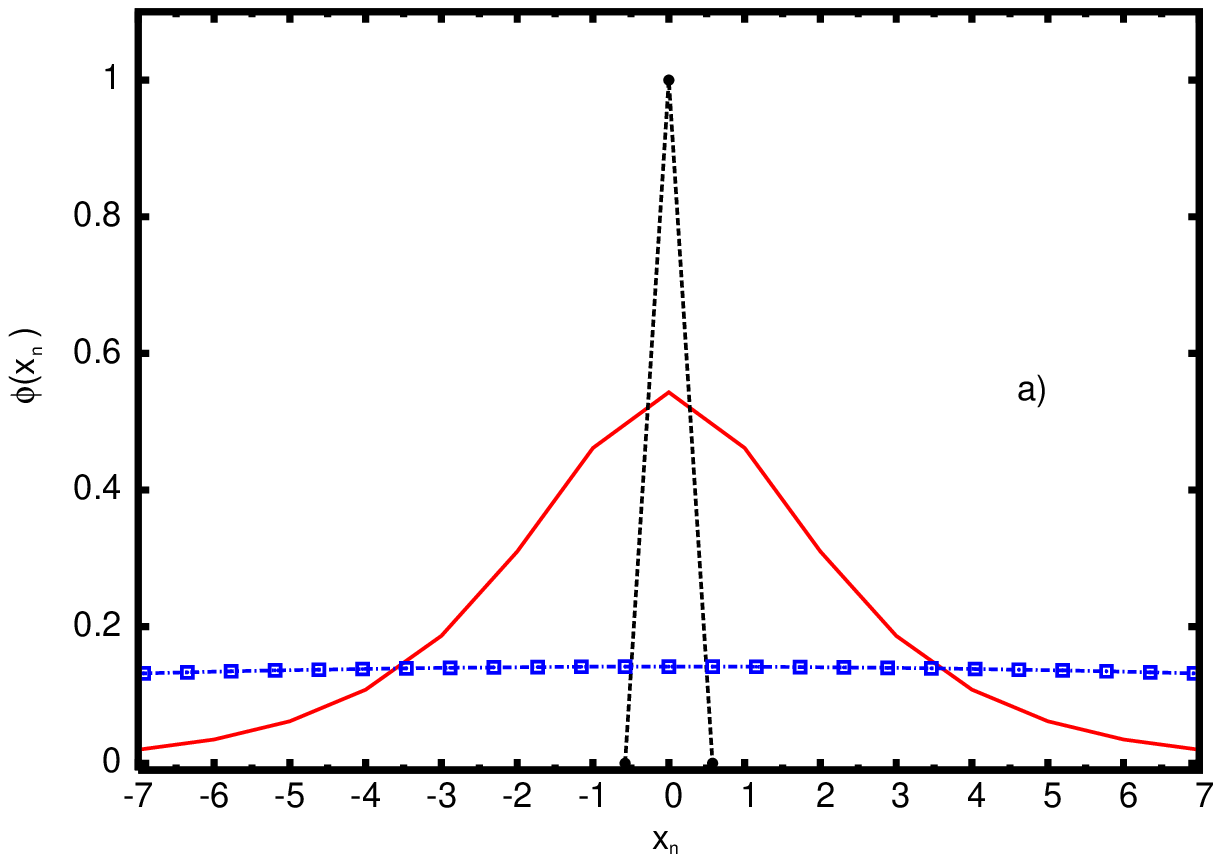} &
    \includegraphics[scale=0.69]{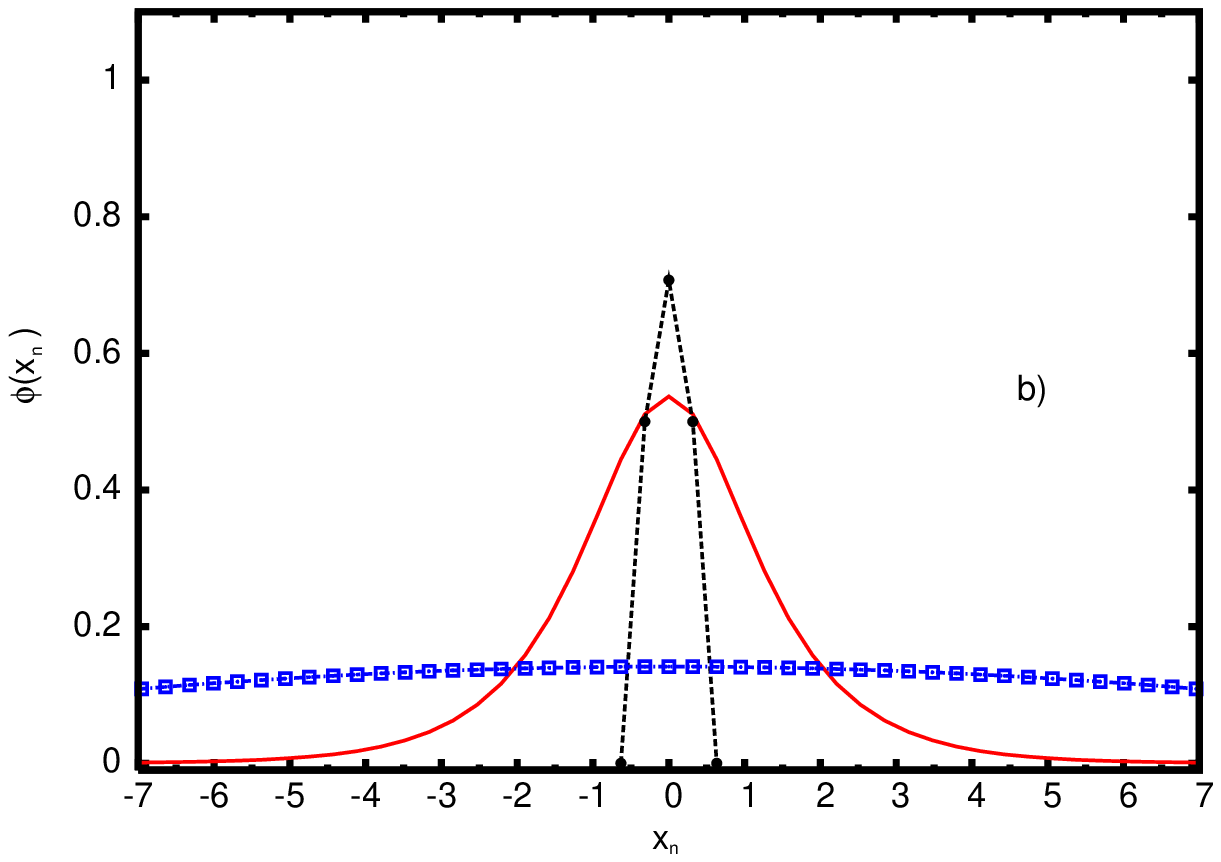}
    \end{tabular}
\caption{(a) Breather profile for $\epsilon=3$ (continuous (red) curve) against the eigenvector (dashed (black) curve) of (\ref{br10A})-(\ref{br10B}) on the  interval $U'$ of length $L'\sim 1.154$. The eigenvector of (\ref{DLap}) in the length $L$ of the chain is represented by the dashed-boxes (blue) curve. Other parameters are $\alpha=1$, $\beta=5$, $\Omega=1$.  (b) Breather profiles for
$\epsilon=10$. Here $L'\sim 1.264$.  Other parameters  $\alpha=1$, $\beta=5$, $\Omega=1$.}
\label{figprofs3_10}
\end{center}
\end{figure}
%%%%%%%%%%%%%%%%%%%%%%%%%%%%%%%%%%%%%%%%%%%%%%%%%%%%%%%%%%%%%%%%%%%%%%%%%%%%%%%%%%%%%%%%%%%%%%%%%%%%%%%%%%%%%%%%%%%
\subsubsection{Numerical results: Quintic nonlinearity $\sigma=2$}
\label{subs3}
Concerning the case of non-cubic nonlinearity ($\sigma\ne1)$, an explicit
estimate from equation (\ref{eqRfv1}) comes out
\begin{eqnarray} \label{focsum2}
 \left[\frac{1}{2\beta}\left(\Omega+\mu_1
-\frac{(2\alpha
N)^{\frac{\sigma}{\sigma-1}}}{(\beta\sigma)^{\frac{1}{\sigma-1}}}
\frac{\sigma-1}{\sigma}\right)\right]^{\frac{1}{\sigma}}<\mathcal{P},\;\;\sigma>1,\;\;N\geq
1,
\end{eqnarray}
with some restriction on the parameters $\sigma$, $\Omega$ and
$\epsilon$ given in Theorem \ref{posFreq} C: \newline
(i) for all
$\Omega>0$ if
\begin{eqnarray}
\label{smeshA}
\epsilon>\frac{(2\alpha N)^{\frac{\sigma}{\sigma-1}}}{(\beta\sigma)^{\frac{1}{\sigma-1}}}\frac{\sigma-1}{\lambda_1\sigma},\;\;\sigma>1,\;\;N\geq 1,
\end{eqnarray}
 and (ii) for all $\epsilon>0$ if
\begin{eqnarray}
\label{sfreqmeshA}
\Omega>\frac{(2\alpha N)^{\frac{\sigma}{\sigma-1}}}{(\beta\sigma)^{\frac{1}{\sigma-1}}}\frac{\sigma-1}{\sigma},\;\;\sigma>1,\;\;N\geq 1.
\end{eqnarray}
In the non-cubic case, the cut-off approximation of Proposition \ref{UL} leads to
\begin{corollary}
\label{corUL}
Let $\sigma>1$, $N=1$. Then the estimate (\ref{focsum2}) is valid with $\mu_1$ replaced by
\newline
A. $\mu_{1,U'}=4\epsilon\sin^2\left(\frac{\pi}{2(m+1)}\right)$ for any $\epsilon>0$.
\newline
B. $\mu_{1,U'}\sim 4\epsilon\sin^2\left(\frac{\pi}{2\sqrt{\epsilon}}\right)$ when $\epsilon>0$ is sufficiently large.
\end{corollary}

The theoretical estimates proposed in section \ref{enot3} for infinite
lattices can also be used. While an infinite lattice cannot be modelled numerically,
the estimates of section \ref{enot3} can serve as alternatives to
those summarized above for the finite lattice. The unspecified
parameter $\nu_{\mathrm{crit}}$ involved in (\ref{res1}),  in the
estimate (\ref{unspec}) and  restrictions
(\ref{mesh})-(\ref{freqmesh}) has been determined by justified
heuristic (and rigorous in the case of ``large'' $\sigma$) arguments
in \cite[Section III, pg. 7]{JFN2009}. For instance it was revealed
that the value $\nu_{\mathrm{crit}}=1$ is valid for all $N\geq 1$
and $\sigma\geq 1$. Furthermore, this value is of very good accuracy
for $N=2$ and excellent for $N=3$. Let us also recall that this
value covers when $\sigma\in\mathbb{N}$, the cases which are of main
physical interest (see also \cite{DorZhouCam08} considering integer
values of $\sigma\geq 2/N$).

For $\nu_{\mathrm{crit}}=1$, Theorem \ref{notrih2} predicts that for
supercritical nonlinearity  $\sigma\geq 2/N$ any breather solution
must have power
\begin{eqnarray}
\label{critroot} \hat{R}^2_{\mathrm{crit}}<\mathcal{P}.
\end{eqnarray}
$\hat{R}_{\mathrm{crit}}$ is the positive root of  the equation
\begin{eqnarray}
 2\alpha N R^2+\beta R^{2\sigma}-\left(\Omega+\frac{4\epsilon N}{2\sigma
 +1}\right)=0.
\end{eqnarray}
Theorem \ref{fintheo} gives explicitly
\begin{eqnarray}
\label{unspec1}
 \left[\frac{1}{2\beta}\left(\Omega+\frac{4\epsilon N}{2\sigma+1}
-\frac{(2\alpha
N)^{\frac{\sigma}{\sigma-1}}}{(\beta\sigma)^{\frac{1}{\sigma-1}}}\frac{\sigma-1}{\sigma}\right)\right]^{\frac{1}{\sigma}}<R^2,\;\;\sigma\geq
2\;\;\mbox{when $N=1$}\;\;\mbox{and}\;\;\sigma> 1\;\;\mbox{when
$N\geq 2$},
\end{eqnarray}
in either  of the cases below:
\newline (i) for all $\Omega>0$ and lattice spacing
satisfying
\begin{eqnarray}
 \label{mesh1}
\epsilon>\frac{(2\alpha
N)^{\frac{\sigma}{\sigma-1}}}{(\beta\sigma)^{\frac{1}{\sigma-1}}}\frac{(\sigma-1)(2\sigma+1)}{4N\sigma},\;\;
\sigma\geq 2\;\;\mbox{when $N=1$}\;\;\mbox{and}\;\;\sigma>
1\;\;\mbox{when $N\geq 2$},
\end{eqnarray}
and \newline (ii) for all $\epsilon>0$ and frequencies
\begin{eqnarray}
 \label{freqmesh1}
\Omega>\frac{(2\alpha
N)^{\frac{\sigma}{\sigma-1}}}{(\beta\sigma)^{\frac{1}{\sigma-1}}}\frac{\sigma-1}{\sigma},\;\;
\sigma\geq 2\;\;\mbox{when $N=1$}\;\;\mbox{and}\;\;\sigma>
1\;\;\mbox{when $N\geq 2$}.
\end{eqnarray}

Additionally other choices of the parameter $\hat{\epsilon}$ in the
Young's inequality trick (see Theorem \ref{posFreq}C), give versions
of the estimates valid with different restrictions on the coupling parameter
$\epsilon$ or the frequency $\Omega$. Together with the choice used
in Theorem \ref{posFreq}C, another interesting one is the  standard
$\hat{\epsilon}=1$ corresponding to the version of (\ref{unspec})
\begin{eqnarray}
\label{Y1unspec1}
 \left[\frac{\sigma}{\sigma\beta+1}\left(\Omega+\frac{4\epsilon N}{2\sigma+1}
-\frac{(\sigma-1)(2\alpha
N)^{\frac{\sigma}{\sigma-1}}}{\sigma}\right)\right]^{\frac{1}{\sigma}}<R^2,\;\;\sigma\geq
2\;\;\mbox{when $N=1$}\;\;\mbox{and}\;\;\sigma> 1\;\;\mbox{when
$N\geq 2$}
\end{eqnarray}
The estimate (\ref{Y1unspec1}) is valid\newline (i) for all
$\Omega>0$ and lattice spacing satisfying
\begin{eqnarray}
 \label{Y1mesh1}
\epsilon>\frac{(2\alpha
N)^{\frac{\sigma}{\sigma-1}}(\sigma-1)(2\sigma+1)}{4N\sigma},\;\;
\sigma\geq 2\;\;\mbox{when $N=1$}\;\;\mbox{and}\;\;\sigma>
1\;\;\mbox{when $N\geq 2$},
\end{eqnarray}
and in the case\newline (ii) for all $\epsilon>0$ and frequencies
\begin{eqnarray}
 \label{Yfreqmesh1}
\Omega>\frac{(2\alpha
N)^{\frac{\sigma}{\sigma-1}}(\sigma-1)}{\sigma},\;\; \sigma\geq
2\;\;\mbox{when $N=1$}\;\;\mbox{and}\;\;\sigma> 1\;\;\mbox{when
$N\geq 2$}.
\end{eqnarray}

%%%%%%%%%%%%%%%%%%%%%%%%%%%%%%%%%%%%%%%%%%%%%%%%%%%%%%%%%%%%%%%%%%%%%%%%%%%%%%%%%%
\begin{figure}[h!]
\begin{center}
    \begin{tabular}{cc}
   \includegraphics[scale=0.69]{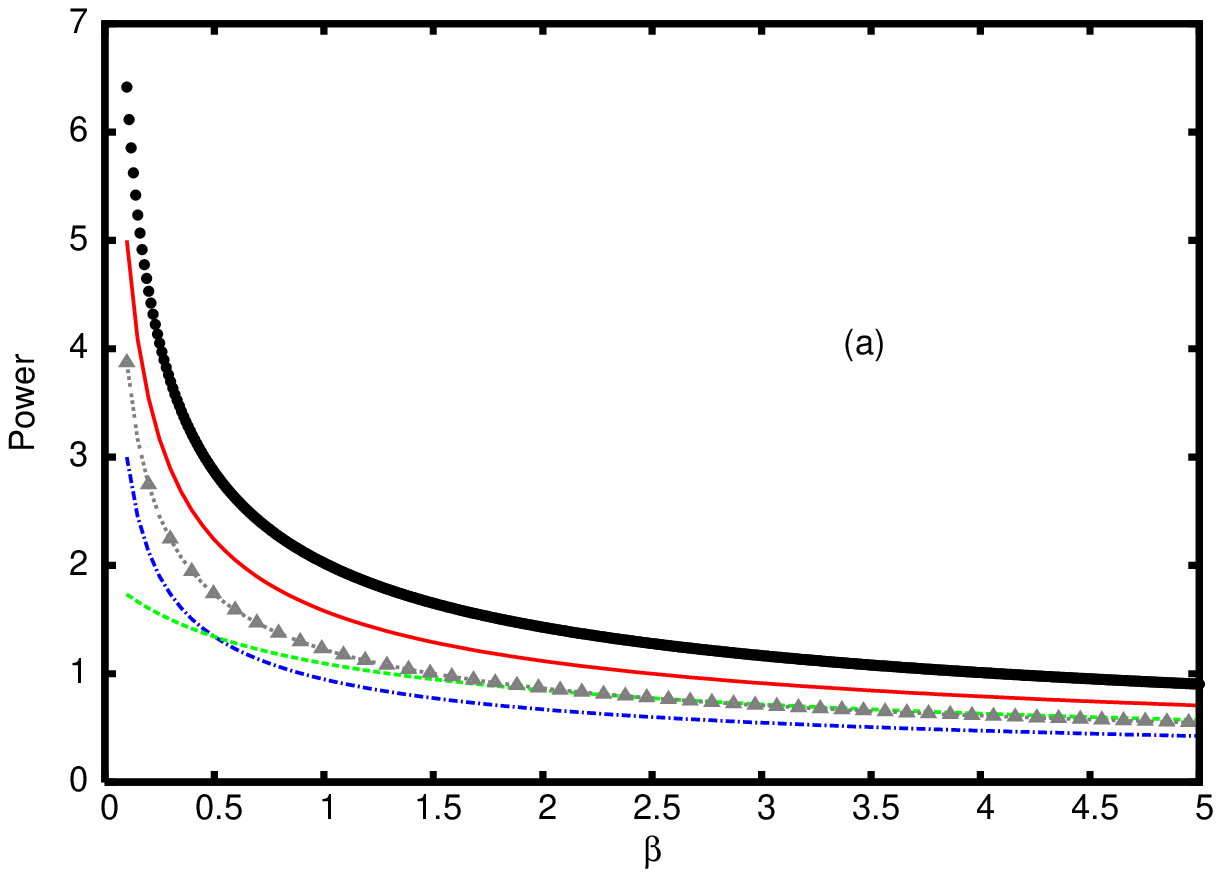}&
  \includegraphics[scale=0.69]{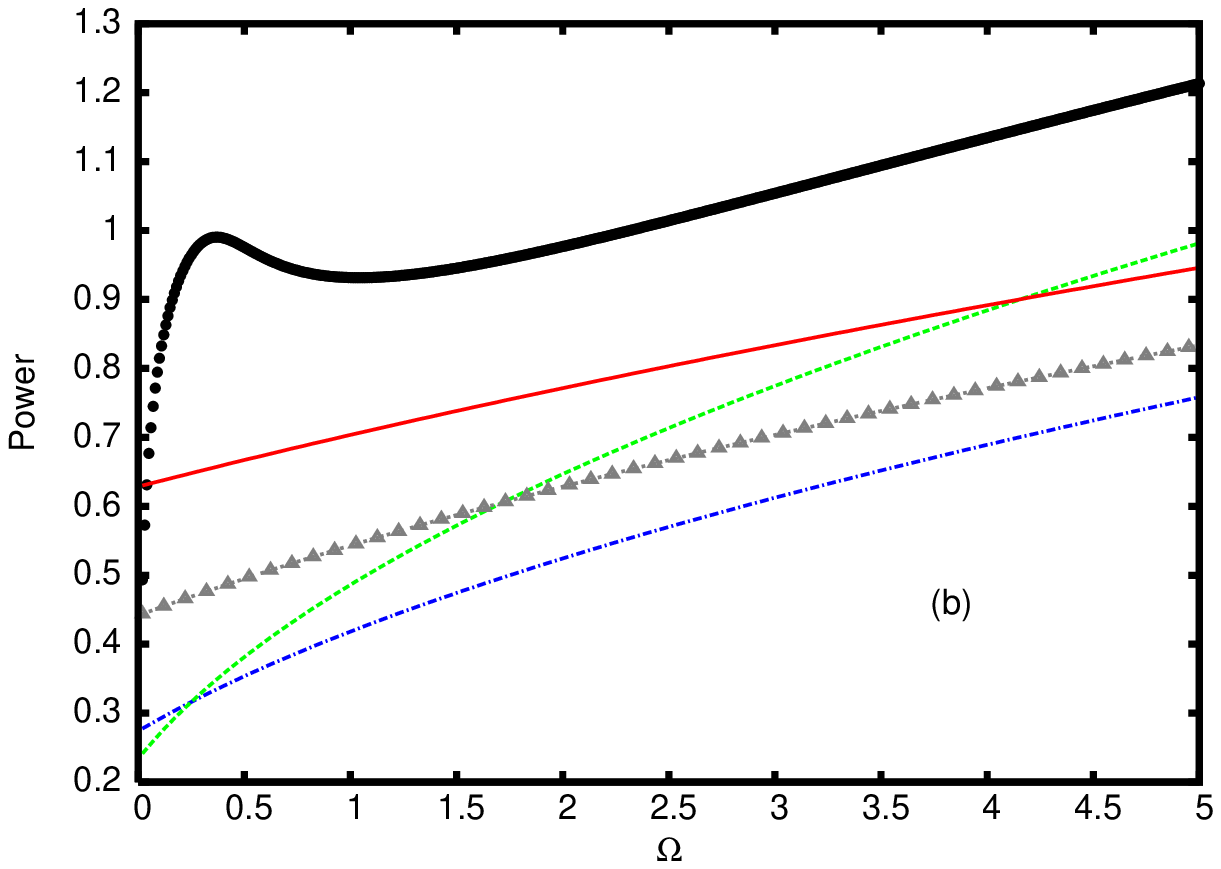}
    \end{tabular}
\caption{(a) Power of breathers versus parameter $\beta$ for
supercritical nonlinearity $\sigma=2$ in the HDNLS system. Symbols
correspond to numerical calculations, the dash-dotted (blue) line represents estimation (\ref{unspec1}) and the dashed line (green) estimation
(\ref{Y1unspec1}). The estimate (\ref{focsum2})-A. of Corollary \ref{corUL} corresponds to
the triangles (grey line), and the estimate (\ref{focsum2})-B. with the continuous (red) line. Parameters: $\alpha=0.01$, $\Omega=\epsilon=1$.
(b) Power versus frequency $\Omega$ for supercritical nonlinearity
$\sigma=2$.  Parameters: $\alpha=0.5$, $\beta=5$ and
$\epsilon=1$. \label{fig3}}
\end{center}
\end{figure}
%%%%%%%%%%%%%%%%%%%%%%%%%%%%%%%%%%%%%%%
Regarding the quintic nonlinearity, more 
specifically, 
we have performed a test of the estimates (\ref{focsum2}), (\ref{unspec1}) and
(\ref{Y1unspec1}) by fixing $\sigma=2$, and $\epsilon=1$. With these
choices,  restrictions (\ref{sfreqmeshA})-(\ref{freqmesh1}) and (\ref{Yfreqmesh1})
reduce to the very simple conditions $\Omega>\alpha^2/\beta$ and
$\Omega>2\alpha^2$. We expect all the estimates to be satisfied as thresholds due to the increased strength of the power nonlinearity absorbing the contribution of the linear part, even in the case of (\ref{focsum2})-B., which is not justified theoretically. In Fig.\ref{fig3}(a) we have plotted the estimate (\ref{focsum2})-A. of Corollary \ref{corUL},
with triangles (grey curve), and its case B. with the continuous (red) curve.  The
dash-dotted (blue) and dashed (green) lines correspond to (\ref{unspec1}) and (\ref{Y1unspec1})  respectively.  The
numerical power (symbols) was obtained  varying $\beta$ for a small
hopping parameter $\alpha=0.01$ and $\Omega=1$. Note that all the
estimates are good, although (\ref{Y1unspec1}) is better than
(\ref{unspec1}) for large $\beta$ while (\ref{unspec1}) behaves
better when $\beta<1/\sigma$.

In Fig. \ref{fig3}(b) we have plotted the breather
power against $\Omega$ choosing
$\beta=5$ and $\alpha=0.5$. Condition (\ref{sfreqmeshA}) is fulfilled for $\Omega>0.05$ and condition (\ref{Yfreqmesh1}) is
fulfilled for $\Omega>0.5$. In the latter region, since $\beta$ is quite
large, the estimate (\ref{Y1unspec1})  behaves clearly
better than (\ref{unspec1}). It is
interesting to realize that (\ref{focsum2}) is worse than
(\ref{Y1unspec1}) for large enough frequencies.

%%%%%%%%%%%%%%%%%%%%%%%%%%%%%%%%%%%%%%%%%%%%%%%%%%%%%%%%%%%%%%%%%%%%%%%%%%%%%%%%%%
%%%%%%%%%%%%%%%%%%%%%%%%%%%%%%%%%%%%%%%%%%%%%%%%%%%%%%%%%%%%%%%%%%%%%%%%%%%%%%%%%%
\subsection{Defocusing case ($\alpha<0,\beta=<0$)
with Dirichlet boundary conditions. Solutions
$\psi_n(t)=e^{-\mathrm{i}\Omega t}\phi_n$, $\Omega>0$.} \label{subB}
In the defocusing case the results on the theoretical estimates are
restricted to frequencies $\Omega>4N\epsilon$. In this case, setting
for convenience $\kappa=-\alpha>0, \lambda=-\beta>0$, the results of
Theorem \ref{negFreq} state that for all $\sigma>0$ the lower bound
for the power of the staggered breathers is given by the positive
root $R_{*,d}$ of the equation
\begin{eqnarray}
\label{eqRav0}
\lambda\chi^{2\sigma}+2\kappa N\chi^2-(\Omega-4\epsilon N)=0,\;\;\sigma>0,\;\;N\geq 1,\;\;\Omega>4\epsilon N,
\end{eqnarray}
and the power of staggered breathers satisfies
\begin{eqnarray}
\label{eqRav1}
 R_{*,d}^2<\mathcal{P}, \;\;\mbox{for all}\;\;\sigma>0,\;\;N\geq 1,\;\;\Omega>4\epsilon N.
\end{eqnarray}

In the defocusing case and \textit{cubic nonlinearity}, the lower bound for the power is
\begin{eqnarray}
\label{decubicRegion2}
\frac{\Omega-4\epsilon N}{2\kappa N +\lambda}<\mathcal{P},\;\;\Omega>4\epsilon N,\;\;N\geq 1\;\;\sigma=1.
\end{eqnarray}
The explicit estimate valid for $\sigma>1$ is
\begin{eqnarray}
\label{defocsum2}
 \left[\frac{1}{2\lambda}\left(\Omega-4\epsilon N
-\frac{(2\kappa N)^{\frac{\sigma}{\sigma-1}}}{(\lambda\sigma)^{\frac{1}{\sigma-1}}}\frac{\sigma-1}{\sigma}\right)\right]^{\frac{1}{\sigma}}<\mathcal{P},\;\;\sigma>1,\;\;N\geq 1,\;\;\Omega>4\epsilon N+\frac{(2\kappa N)^{\frac{\sigma}{\sigma-1}}}{(\lambda\sigma)^{\frac{1}{\sigma-1}}}\frac{\sigma-1}{\sigma}.
\end{eqnarray}
%%%%%%%%%%%%%%%%%%%%%%%%%%%%%%%%%%%%%%%%%%%5555

The results of the numerical tests in the defocusing case are
similar to those of the focusing case and  can be summarized in the
following points:
\begin{itemize} \item The theoretical estimates are always below the
numerical power and approximate quite well the nonlinear part of the
contribution to the power. \item The lower bound (\ref{eqRav1}) is
always above the explicit estimate (\ref{defocsum2}) \item Estimate
(\ref{defocsum2}) behaves better for small values of the hopping
parameter $\alpha$ and  large exponents $\sigma$.
\end{itemize}
These observations are corroborated by the results of
Fig.\ref{fig4}. Squares and
the upper continuous curve correspond respectively to the numerical
power and estimate (\ref{eqRav1}) for $\beta=-1$ and a hopping
parameter $\alpha=-0.5$. The estimate becomes much closer the real
power fixing $\beta=-5$ and $\alpha=-0.01$ (see pluses and the lower
continuous curve).

%%%%%%%%%%%%%%%%%%%%%%%%%%%%%%%%%%%%%%%%%%%%%%%%%%%%%%%%%%%%%%%%%%%%%%%%%%%%%%%%%%
\begin{figure}[h!]
\begin{center}
    \begin{tabular}{cc}
    \includegraphics[scale=0.7,angle=-0]{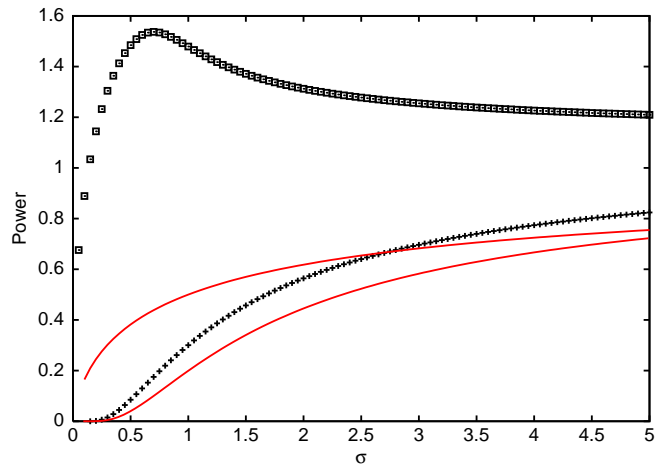}
    \end{tabular}
\caption{Power of breathers versus parameter $\sigma$ in the defocusing
case. Squares (pluses) correspond to the numerical power found for
 $\beta=-1,\alpha=-0.5$  ($\beta=-5,\alpha=-0.01$) while the
upper (lower) continuous line represents estimate (\ref{defocsum2}).
Other parameters: $\Omega=-2, \epsilon=0.25$. \label{fig4}}
\end{center}
\end{figure}
%%%%%%%%%%%%%%%%%%%%%%%%%%%%%%%%%%%%%%%

\section{Conclusions}

In the present work, we generalized the considerations of
energy thresholds in the setting of a DNLS model with
generalized nonlinear (Hamiltonian) hopping terms.
Different types of bounds were provided for the power both
for finite and for infinite lattices, by using appropriate
estimates for the linear coupling and nonlinear hopping terms.
A fixed point method establishing the contractivity of
an appropriately defined operator was also used to establish
that for a given parameter set, there is a critical power,
below which it is not possible to sustain such nonlinear
waveforms. Finally, some dimension-dependent estimates
were given based on the interpolation inequality of
the Gagliardo-Nirenberg type, in a spirit similar to the
work of \cite{Wein99}.

Further improvements of the main theory have been considered and proved, appreciating
the interplay of the nonlinear and linear term contributions
within the true solitary wave solutions, taking into account
their spatial localization.
The obtained bounds were tested
numerically and in all the cases where the theory was expected to be
applicable, it was found that the
numerical solutions satisfy the predicted norm inequalities.
This aspect also provides details on the parameter regimes
(weak linear coupling) which tend
to saturate the corresponding theoretically obtained bounds.

We are leaving as an interesting open direction for a future work, to examine the behavior of the energy bounds when the size of the lattice is varied. This question is taking into account the effect of the transition from finite to infinite lattices, on the localization properties of the solutions. This task could be based on a generalization and use of the machinery developed in  \cite{PP}, as well as,  of the relevant localization estimates. Such a generalization could be of particular interest, in the case of multidimensional lattices.

It would be also interesting and relevant to examine how
corresponding bounds can be generalized to other classes of
models, including ones of the nonlinear Klein-Gordon or
FPU type (or mixed ones), incorporating different types of onsite and
intersite nonlinearities.  Especially useful, albeit
arguably more difficult, to extend the main strategy to continuous models.
Such tasks will be considered in future publications.

%%%%%%%%%%%%%%%%5555555
%\section*{References}

\bibliographystyle{amsplain}

\end{document}